%% file: paper.tex
\documentclass[sigconf]{acmart}

\usepackage{xcolor}

\usepackage{soul}
\sethlcolor{yellow}

\usepackage{color, colortbl}
\usepackage[utf8]{inputenc}
\usepackage{booktabs}
\usepackage[table]{xcolor} 
\usepackage{pgf} 
\usepackage{collcell} 
\usepackage{subcaption}
\usepackage{tabularx}
\usepackage{balance}
\usepackage{longtable}
\definecolor{LightGray}{rgb}{0.88,0.87,0.88}

\usepackage{pgfplots}
\pgfplotsset{compat=1.17} 

\usepackage{listings}
\usepackage{newunicodechar}

\usepackage{listings}
\usepackage{newunicodechar}

\DeclareUnicodeCharacter{00A0}{ }      
\DeclareUnicodeCharacter{00A4}{¤}      
\DeclareUnicodeCharacter{00AD}{-}      
\DeclareUnicodeCharacter{200B}{}       
\DeclareUnicodeCharacter{200E}{}       
\DeclareUnicodeCharacter{200F}{}       
\DeclareUnicodeCharacter{2028}{\\}     
\DeclareUnicodeCharacter{202F}{ }      
\DeclareUnicodeCharacter{2060}{}       
\DeclareUnicodeCharacter{FE0F}{}       
\DeclareUnicodeCharacter{FEFF}{}       

\usepackage[T1]{fontenc}

\usepackage{listings}
\usepackage{booktabs,multirow}
\usepackage{minted}

\usepackage{booktabs}
\usepackage{placeins}
\usepackage{tikz}
\usetikzlibrary{
  shapes.geometric,
  arrows.meta,
  positioning,
  shadows
}

\usepackage{booktabs}
\usepackage{placeins}
\usepackage{tikz}
\usetikzlibrary{
  shapes.geometric,
  arrows.meta,
  positioning,
  shadows
}

\usepackage{enumitem}

\usepackage{tikz}
\usetikzlibrary{
  shapes.geometric,
  arrows.meta,
  positioning,
  shadows
}

\usepackage{tikz}
\usetikzlibrary{
  shapes.geometric,
  arrows.meta,
  positioning,
  shadows,
  matrix,
  calc          
}
\tikzset{
  process/.style={
    rectangle,
    rounded corners,
    draw=blue!80,
    fill=blue!10,
    thick,
    text width=3cm,
    minimum height=0.9cm,
    align=center,
    drop shadow
  },
  data/.style={
    trapezium,
    trapezium left angle=70,
    trapezium right angle=110,
    draw=green!70,
    fill=green!10,
    thick,
    text width=3.5cm,       
    minimum height=0.9cm,
    align=center,
    drop shadow
  },
  fusion/.style={
    diamond,
    draw=red!80,
    fill=red!10,
    thick,
    text width=2.8cm,
    minimum height=1cm,
    align=center,
    drop shadow
  },
  arrow/.style={
    -{Latex[length=2mm,width=1.5mm]},
    thick
  }
}

\usepackage{xspace,soul}

\newcommand{\eg}{e.g.\@\xspace}

\newcommand{\etal}{et~al.\@\xspace}
\newcommand{\US}{U.\kern0.5ptS.\@\xspace}

 \newcommand{\myparagraph}[1]{\smallskip\noindent\textbf{\textit{#1}.}\hspace{3pt}}
 \newcommand{\name}{\texttt{PrivAudit}\@\xspace}
  \newcommand{\subject}{\texttt{CCPA\allowbreak-Subject}\@\xspace}
   \newcommand{\notsubject}{\texttt{CCPA\allowbreak-Not\allowbreak-Subject}\@\xspace}
     \newcommand{\totalwebsitesbeforefiltering}{998\@\xspace}
  \newcommand{\totalwebsites}{915\@\xspace}

    \newcommand{\privacypolicies}{933\@\xspace}

\newcommand*{\minval}{0.0}
\newcommand{\gradientcell}[6]{
    \ifdimcomp{#1pt}{>}{#3 pt}{#1}{%
        \ifdimcomp{#1pt}{<}{#2 pt}{#1}{%
            \pgfmathparse{int(round(100*(#1/(#3-#2))-(\minval*(100/(#3-#2)))))}%
            \xdef\tempa{\pgfmathresult}%
            \cellcolor{#5!\tempa!#4!#6} #1%
    }}\ignorespaces%
}

\newcommand{\gradientcellwtext}[7]{ 
    \ifdimcomp{#1pt}{>}{#3 pt}{#7}{%
        \ifdimcomp{#1pt}{<}{#2 pt}{#7}{%
            \pgfmathparse{int(round(100*(#1/(#3-#2))-(\minval*(100/(#3-#2)))))}%
            \xdef\tempa{\pgfmathresult}%
            \cellcolor{#5!\tempa!#4!#6} #7%
    }}
}

\usepackage[most]{tcolorbox}
\usepackage{xcolor}

\newcounter{takeaway}

\newcommand{\takeaway}[1]{%
  \refstepcounter{takeaway}%
  \begin{tcolorbox}[
    breakable,
    colback=gray!12,
    colframe=gray!45,
    boxrule=0.4pt,
    arc=2pt,
    left=6pt,right=6pt,top=5pt,bottom=5pt,
    width=\columnwidth,
    before skip=6pt,
    after skip=6pt
  ]
  \textbf{Takeaway \thetakeaway.} #1
  \end{tcolorbox}
}

\copyrightyear{2026}
\acmYear{2026}
\setcopyright{cc}
\setcctype{by}
\acmConference[CCS '26]{Proceedings of the 2026 ACM SIGSAC Conference on Computer and Communications Security}{November 15--19, 2026}{The Hague, Netherlands}
\acmBooktitle{Proceedings of the 2026 ACM SIGSAC Conference on Computer and Communications Security (CCS '26), November 15--19, 2026, The Hague, Netherlands}
\acmDOI{10.1145/3830454.3846679}
\acmISBN{979-8-4007-2871-6/2026/11}

\makeatletter
\def\@copyrightpermission{%
  \begin{minipage}{0.3\columnwidth}%
    \href{https://creativecommons.org/licenses/by/4.0/}%
      {\includegraphics[width=0.9\textwidth]{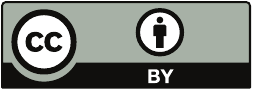}}%
  \end{minipage}\hfill
  \begin{minipage}{0.68\columnwidth}%
    \href{https://creativecommons.org/licenses/by/4.0/}%
      {This work is licensed under a Creative Commons Attribution 4.0 International License.}%
  \end{minipage}%
}
\makeatother

\newlength{\qswidth}
\newlength{\nswidth}

\setcitestyle{notesep={~}}

\begin{document}

\title{PrivAudit: A Dual-Lens Auditing Framework for Website Privacy Practices under the CCPA}

\author{Mohamed Moustafa Dawoud}
\affiliation{%
  \institution{University of California, Santa Cruz}
  \city{Santa Cruz}
  \state{California}
  \country{USA}}
\email{mdawoud@ucsc.edu}

\author{Riya Aggarwal}
\affiliation{%
  \institution{University of California, Santa Cruz}
  \city{Santa Cruz}
  \state{California}
  \country{USA}}
\email{raggarw6@ucsc.edu}

\author{Likith Rahul Krishnamurthy}
\affiliation{%
  \institution{University of California, Santa Cruz}
  \city{Santa Cruz}
  \state{California}
  \country{USA}}
\email{likrishn@ucsc.edu}

\author{Ram Sundara Raman}
\affiliation{%
  \institution{University of California, Santa Cruz}
  \city{Santa Cruz}
  \state{California}
  \country{USA}}
\email{rsundar2@ucsc.edu}

\renewcommand{\shortauthors}{Dawoud et al.}

\input{0-abstract}



\maketitle

\input{1-introduction}

\input{2-background}

\input{3-methodology}

\input{4-results}
\input{5-discussion}

\begin{acks}
We thank the anonymous reviewers for their constructive feedback. We are grateful to Sabrina Ross for her expert insights and review of this work.
We also thank Reethika Ramesh and Hieu Le for their valuable feedback, as well as Isa Abello, Kenneth Lai, Kelvin Chan, and other members of our research group for their insights.
\end{acks}

\bibliographystyle{ACM-Reference-Format}
\bibliography{paper}
\input{99-appendix}
\end{document}

%% file: 0-abstract.tex
\begin{abstract}
Five years after the enforcement of the California Consumer Privacy Act (CCPA), understanding how website privacy practices evolve at scale in response to regulation remains a key challenge for both researchers and regulators. Prior work and regulatory efforts have focused on manual and case-specific enforcement, but there remain no scalable approaches to systematically audit two key user-facing facets of websites that are crucial signals for the CCPA: privacy disclosures and front-end user tracking behavior. 

In this paper, we present \name, an automated auditing framework that adopts a dual-lens approach to capture: (1) privacy disclosures through large language model-based analysis of privacy policies grounded in CCPA provisions, and (2) user-observable data collection behavior through automated browser measurements of cookie writes under diverse privacy configurations. We apply \name to 998 websites and report two broad findings. The law is associated with stronger privacy disclosures: CCPA-subject policies are more likely to disclose opt-out mechanisms, data-sharing practices, and user rights. On the other hand, cookie-based tracking remains pervasive, with both CCPA-subject and not-subject websites setting a total of 6,392 targeting cookies, 49\% of which are third-party writes. Moreover, cookies show limited-to-moderate responsiveness to privacy signals and consent choices, even when websites claim to honor them in their disclosures.

Our results highlight the need for multi-layered and scalable auditing approaches that combine policy analysis with behavioral evidence. \name can support these auditing workflows at scale by generating actionable signals and patterns for further manual review. We open-source \name and are engaging with regulators to support auditing in practice.
\end{abstract}

%% file: 1-introduction.tex
\section{Introduction}
The California Consumer Privacy Act (CCPA), as amended by the California Privacy Rights Act (CPRA), is the first comprehensive consumer data privacy law in the United States~\cite{ccpa}. The CCPA grants California residents the right to know what personal information businesses collect, to request its deletion, and to opt out of its sale or sharing with third parties. Unlike the opt-in model of the EU's General Data Protection Regulation (GDPR), which requires businesses to obtain consent \textit{before} collecting data~\cite{gdpr2016regulation}, the CCPA allows businesses to collect data by default and places the burden on users to opt out. To exercise this right, users can submit requests through website interfaces or send automated browser signals such as the Global Privacy Control (GPC), which CCPA-covered businesses must honor as an opt-out preference signal~\cite{gpc_spec}. More than five years after the CCPA's enforcement, and as more than 20 U.S.\ states adopt similar privacy frameworks~\cite{iapp-us-privacy-laws-2025,co_cpa_scope,va_cdpa_rights,ut_ucpa_rights}, understanding how websites' privacy practices have evolved under the law, and how they can be audited at scale, remains an open challenge for both researchers and regulators.

Despite the growing importance of the CCPA, both regulatory enforcement and academic measurement remain limited in scope. Public enforcement actions by the California Attorney General have largely proceeded case by case, focusing on specific companies and documented practices such as undisclosed tracking and ineffective opt-out mechanisms~\cite{sephora_ccpa_2022,tractor_supply_cppa,honda_cppa,healthline_ccpa_2025,ccpa_enforcement_examples}. This case-specific model is essential for legal enforcement, but difficult to scale to broad audits because regulators must manually evaluate potential violations~\cite{cppa_annual_report}.
Academic research has similarly examined isolated, point-in-time compliance questions, including opt-out link deployment~\cite{tran2024measuring,van2022setting}, dark patterns in consent interfaces~\cite{tran2025dark,o2021clear,mazumdar2023current}, and downstream ad-tech behavior~\cite{aziz2024johnny,Liu_2024}.  Meanwhile, the more extensive GDPR literature has shown that tracking mechanisms can undermine user choice~\cite{matte2020cookie,Degeling_2019,bollinger2022automating,nouwens2020dark}. The CCPA's opt-out model~\cite{charatan2024twosteps,o2021clear} has received less empirical scrutiny~\cite{birrell2024sok}, particularly around the user-facing privacy practices. 

Although this body of work provides a valuable foundation of website compliance with the CCPA, two important and inter-related gaps remain regarding privacy practices of websites: (1) the lack of scalable and practical methods to systematically analyze privacy policy disclosures under the CCPA, and (2) limited understanding of cookie writing practices for CCPA-subject and not-subject websites. Importantly, these two elements together provide \textit{externally observable and complementary user-facing signals of website privacy practices} without privileged access, making them ideal for the \textit{auditing} needs of California regulators~\cite{cppa_strategic_plan}, which involves multi-layered assessments of privacy practices and trends at scale across numerous sectors.
This dual-lens view is also increasingly relevant to regulatory practice. Privacy policies are the primary interface through which websites communicate data practices, rights, and opt-out mechanisms. Cookie writes, in turn, often represent the first observable step in user data collection and can reveal whether tracking persists under privacy signals such as GPC. These dimensions are legally and practically connected. The CCPA defines personal information to include online identifiers such as cookies, pixel tags, and device identifiers (Cal.\ Civ.\ Code \S1798.140\allowbreak(aj)), and the CPRA regulates both the ``sale'' (\S1798.140\allowbreak(ad)) and the ``sharing'' for cross-context behavioral advertising (\S1798.140\allowbreak(ah)) of such information, neither of which requires a monetary exchange; the CPRA correspondingly expanded the consumer's opt-out right from the ``sale'' to the ``sale or sharing'' of personal information (\S1798.120\allowbreak(a))~\cite{cpra_statute}. A third-party script that writes a cookie to store a consumer's identifier can thus initiate the ``sale'' or ``sharing'' the statute regulates, making front-end cookie behavior an observable signal of CCPA-regulated conduct. Recent enforcement actions illustrate the importance of both dimensions~\cite{sephora_ccpa_2022,healthline_ccpa_2025,disney_settlement,tractor_supply_cppa}. For example, Sephora paid \$1.2M in a settlement involving undisclosed third-party tracking technologies, including cookies, and GPC non-compliance~\cite{sephora_ccpa_2022}. In another example, Healthline paid \$1.55M in a case related to third-party sharing of health-related browsing data through cookies and pixels after opt-out, as well as cookie consent banners that falsely claimed to disable tracking~\cite{healthline_ccpa_2025}. Yet such investigations remain case-specific and resource-intensive, motivating scalable auditing methods that can surface disclosure and tracking signals across many websites. 

To address this gap, we develop \textbf{\name}, a scalable dual-lens auditing framework for user-facing privacy practices under the CCPA. \name combines two complementary analyses for surfacing externally observable audit-relevant patterns. First, we develop a \textit{privacy policy audit} that leverages expert-guided large language models (LLMs) to evaluate privacy disclosures against CCPA-grounded rubric dimensions validated through review by a privacy-law expert. We address key challenges in scalable auditing by automating privacy-policy extraction and validating LLM outputs against a human-labeled ground truth (97\% agreement), with a Monte Carlo analysis confirming our findings are robust to LLM error.  Second, we implement a browser-based \textit{cookie audit} that measures first-visit tracking behavior under six different privacy configurations, including GPC, third-party cookie blocking, and consent interactions.  We synthesize multiple datasets to categorize cookie functionality and develop a script-based attribution approach to identify third-party cookies. We ultimately link these two views at the website level to detect disclosure--behavior gaps that prior single-lens approaches cannot detect~\cite{tran2025dark,tran2024measuring,aziz2024johnny,van2022setting}.

We apply \name to study \totalwebsitesbeforefiltering websites (\privacypolicies privacy policies) with known CCPA subjectivity, comparing 602 \subject websites against 396 \notsubject ones. Because the CCPA imposes disclosure and opt-out obligations only on businesses meeting its applicability thresholds, we use \notsubject websites as a descriptive comparison group. We also apply \name to a disjoint stratified sample of 1,000 popular websites from Tranco~\cite{Tranco} without pre-established subjectivity labels as a case study.

We report three main results. First, \textit{\subject websites provide stronger privacy disclosures than \notsubject websites}. They are significantly more likely to describe opt-out mechanisms (77\% vs.\ 57\%), rights to access (85\% vs.\ 68\%) and delete data (86\% vs.\ 69\%), and support for GPC (29\% vs.\ 13\%). 
However, only a small fraction of policies provide detailed disclosures about specific practices, such as cookie consent behavior or honoring  privacy signals. Notably, 43\% of \notsubject websites also reference the CCPA in their policies, suggesting a spillover effect of the law.

Second, \textit{stronger disclosures do not correspond to reduced third-party tracking}. Cookie writes remain widespread across both \subject and \notsubject websites, with 6,392 Targeting cookies in total, almost half of which involve third-party writes. Many third-party cookies are set by scripts operating in the first-party context, and just 10 scripts from large advertising providers account for more than half of all tracking cookies. Privacy signals reduce but do not eliminate tracking: among \subject websites, GPC reduces total tracking cookies by 55\% and third-party tracking cookies by only 41\%. Consent mechanisms are often absent~\cite{habib2022okay}, and when present, user choices have only a limited effect on cookies. Although the CCPA does not explicitly regulate cookie writes themselves, cookies serve as the primary front-end mechanism through which third-party data collection is initiated, and third-party tracking persisting beyond user privacy signals serves as an important auditing signal.

Third, \textit{linking disclosures to cookie behavior reveals audit-relevant gaps}. Among \subject websites whose policies claim to honor GPC, 39\% show no reduction in Targeting cookies when GPC is sent, and more than half reduce Targeting cookies by less than 20\%. The gap is wider for third-party Targeting cookies specifically: 66\% show less than a 20\% reduction and 47\% show no reduction at all. Among \subject websites that state they do not sell personal data, more than half still deploy third-party Targeting cookies at initial load. These findings also generalize across the popular websites without known subjectivity. 
These patterns alone are not evidence of legal violations, but they provide scalable auditing signals for deeper manual review.

These findings underscore the value of multi-lens auditing approaches that combine policy analysis with behavioral evidence. Complementing prior work that has focused on isolated studies of compliance, dark patterns, and data sales~\cite{aziz2024johnny,tran2024measuring, tran2025dark}, \name derives scalable auditing signals that help identify websites, sectors, and practices that may warrant deeper investigation and follow-up enforcement. Although our study focuses on the CCPA, \name is extensible to other state privacy laws that adopt similar disclosure, opt-out, and universal opt-out signal requirements~\cite{iapp-us-privacy-laws-2025}. We have open-sourced \name and all datasets~\footnote{\url{https://github.com/r-andlab/PrivAudit}}, and are engaging with the California regulatory community to explore how it can support auditing in practice.

%% file: 2-background.tex
\section{Background and Related Work}
\label{sec:background}
\myparagraph{The California Consumer Privacy Act (CCPA)}
The CCPA, along with its amendment CPRA, which we jointly refer to as just ``CCPA'', established one of the first comprehensive data privacy frameworks in the United States~\cite{ccpa}. Unlike the GDPR’s strict opt-in regime~\cite{gdpr2016regulation}, the CCPA adopts an opt-out model: businesses may collect personal information by default but must provide clear mechanisms for consumers to opt out of the sale or sharing of their data. Also unlike the GDPR, the CCPA does not bind all businesses and service providers, but only those who (1) do business in California (2) with Californian residents and (3) either (i) buy, sell, or share the personal information of at least 100{,}000 consumers or households, or (ii) had a gross annual revenue of at least US\$25 million, or (iii) generate at least 50\% of their annual revenue from selling or sharing personal information. The CCPA's enactment has spurred a wave of similar legislation across other U.S.\ states~\cite{va_cdpa_rights,co_cpa_scope,ut_ucpa_rights}.

\myparagraph{Prior Empirical Research on CCPA} Empirical work on the CCPA has examined opt-out mechanisms, usability, and policy text, each through a different lens. Studies of opt-out deployment found that many websites bury or geofence opt-out links~\cite{tran2024measuring}, that only $\sim$2\% of 500{,}000 sites displayed the required ``Do Not Sell'' link~\cite{van2022setting}, and that dark patterns in opt-out interfaces suppress user completion rates~\cite{o2021clear,mazumdar2023current}. More recent work documents that these issues persist under the CPRA: Tran~\etal~\cite{tran2025dark} systematically exercised the full opt-out process on 330 websites, finding that nearly 30\% of requests fail~\cite{siebel2022impact}. At the policy-text level, studies have found ambiguity in CCPA-mandated disclosures and internal contradictions in overlapping privacy policies~\cite{chen2021fighting,xian2025layered}. Broader surveys and empirical work further document gaps in tracker disclosure and the difficulty of privacy-compliant implementation~\cite{birrell2024sok,kafle2024understanding,horstmann2025sorry,baik2020data}. 

Our work differs from prior studies in two respects. First, prior CCPA studies generally examine a single compliance dimension: the deployment and readability of ``Do Not Sell'' links~\cite{tran2024measuring,van2022setting}, dark patterns in the opt-out process~\cite{tran2025dark,o2021clear}, or policy text~\cite{chen2021fighting}. In contrast, \name is the first to jointly audit policy disclosures and front-end cookie behavior at the website level, using potential mismatches between them as an audit signal. Examples include a site whose policy states that it does not sell personal data but sets third-party Targeting cookies on page load, or one that claims to honor GPC but exhibits no corresponding reduction in tracking when the signal is sent (\S\ref{sec:policy-behavior-results}). Second, although we reuse the corpus of Tran~\etal~\cite{tran2024measuring} for comparability of the \subject versus \notsubject websites, they audit only the \emph{opt-out mechanism}---whether opt-out links are deployed, readable, and free of dark patterns---whereas we evaluate the \emph{content} of policy disclosures against specific CCPA provisions and measure the \emph{cookie-writing behavior} those policies describe. 


A parallel line of work examines whether opt-out signals are honored downstream in the advertising ecosystem. Aziz~\etal~\cite{aziz2024johnny} audited the IAB CCPA Compliance Framework and found that 90\% of advertising and analytics domains never read the opt-out API. Liu~\etal~\cite{Liu_2024} complemented this by auditing advertisers' real-time bidding behavior, finding that several major ad platforms continue collecting data after opt-out. Technical audits of GPC specifically have found that fewer than half of websites honor the signal across all implemented consent strings~\cite{hausladenwebsites}, that users understand and would enable GPC but backend compliance remains low~\cite{zimmeck2023usability}, and that GPC has minimal effect on app-level ad tracking on Android~\cite{zimmeck2024ad}. Bryson~\etal~\cite{bryson2025characterizing} further documented gaps in how platforms communicate data practices through ad transparency systems. These studies assess \textit{CCPA compliance} by inspecting consent strings such as the US Privacy String and GPP String to determine whether websites update their opt-out status when privacy signals are sent~\cite{hausladenwebsites,zimmeck2023usability,zhang2024cschecker}. However, correct propagation of opt-out signals does not guarantee that data collection actually stops. For example, third-party cookies can persist even after consent rejection~\cite{rasaii2025intractable}, and enforcement actions have penalized businesses whose tracking continued despite opt-out signals and consent interfaces~\cite{tractor_supply_cppa}. Our work complements this body of work by measuring a different layer: the \textit{cookies and tracking scripts} that websites deploy. By testing six privacy configurations---including GPC, Do Not Track (DNT), third-party cookie blocking, ad blocking, and consent banner interaction---and comparing behavior across \subject and \notsubject websites, \name captures whether front-end data collection practices change in response to a variety of user privacy choices.

\input{figures/workflow}
\myparagraph{Advances in Privacy Policy Analysis}
Recent work has leveraged LLMs and natural language processing (NLP) to automate privacy policy analysis, including summarizing and classifying policies~\cite{tang2023policygpt,woodring2024enhancing,salvi2024privacychat,freiberger2025you}, detecting ambiguity and contradictions~\cite{grundler2024detecting,dewri2025interpretation}, evaluating clauses against statutory requirements~\cite{xie2025evaluating,rodriguez2024large,alghamdi2024through,mori2025evaluating}, and tracking how disclosures evolved longitudinally in response to GDPR and CCPA~\cite{wagner2023privacy,hosseini2024bilingual}. Other work has automated policy collection and analysis at scale~\cite{huang2024analyzing} and found that automated policy generators systematically fail to comply with major privacy laws~\cite{pan2024trap}.
Most relevant to our work are systems that compare policy claims against observed behavior. PoliCheck~\cite{andow2020actions} links Android app policies to actual data flows, OVRSeen~\cite{trimananda2022ovrseen} audits VR network traffic against policy statements, PoliGraph~\cite{cui2023poligraph} maps policy text to specific data practices via knowledge graphs, and DiffAudit~\cite{figueira2024diffaudit} uses GPT-4 to compare data flows across user age groups under COPPA and CCPA. Other tools link policy text to app or extension behavior~\cite{ciaramella2022leveraging,andow2019policylint,bui2023detection,ali2023honesty,li2024we,cui2024understanding,pan2024new}. Our work differs from these systems in two ways: we use a \textit{CCPA-specific} rubric mapped to opt-out-based statutory provisions rather than generic privacy practice ontologies, and jointly analyze policy disclosures with observed cookie behavior to surface audit-relevant signals.

\myparagraph{Auditing Cookie-Based Tracking}
Cookie-based tracking is part of a broader ecosystem that includes fingerprinting, evercookies, cookie synchronization, and widespread identifier sharing by a small set of dominant entities~\cite{roesner2012detecting,acar2014web,eckersley2010unique,englehardt2016online,cahn2016empirical,papadopoulos2019cookie}. Recent work shows persistent gaps between consent and tracking: popular sites may offer reject buttons yet ignore rejection~\cite{bouhoula2024automated}, declared consent signals often diverge from actual tracking under the IAB Europe TCF~\cite{smith2024study}, and tracking can persist through respawning, cloaked domains, and incorrect consent data~\cite{fouad2022my,fouad2024devil,zhang2024cschecker}. Finally, third-party tracking can expose sensitive health information beyond what traditional cookie analyses capture~\cite{zeng2025measuring}.

Parallel user studies consistently find that consent interface design shapes tracking outcomes more than user intent.  Dark patterns in cookie banners increase acceptance rates and make rejecting tracking difficult~\cite{nouwens2025cross,nouwens2020dark,Machuletz_2020}, while deployed banners systematically guide users toward acceptance~\cite{habib2022okay,tang2025navigating,Degeling_2019,matte2020cookie,bollinger2022automating,utz2019informed}. Subsequent work has identified additional barriers: standard cookie category labels are poorly understood by users~\cite{jiwani2024crumbling}, automated consent tools vary in effectiveness~\cite{demir2024large}, dark patterns may give rise to legal redress~\cite{gunawan2022redress}, and consent mechanisms are increasingly tied to payment alternatives~\cite{morel2022your,stenwreth2024or,rasaii2025intractable,bouma2023us,habib2022evaluating}.

While prior work on cookies largely focuses on GDPR's opt-in regime, cookie-setting behavior under opt-out models like the CCPA remains underexplored. Our work complements this literature by auditing front-end, user-facing privacy behavior: we measure initial cookie writes under six privacy configurations and jointly analyze these measurements with CCPA-grounded policy disclosures.

%% file: figures/workflow.tex
\begin{figure*}[!t]
    \centering
    \includegraphics[width=1.95\columnwidth]{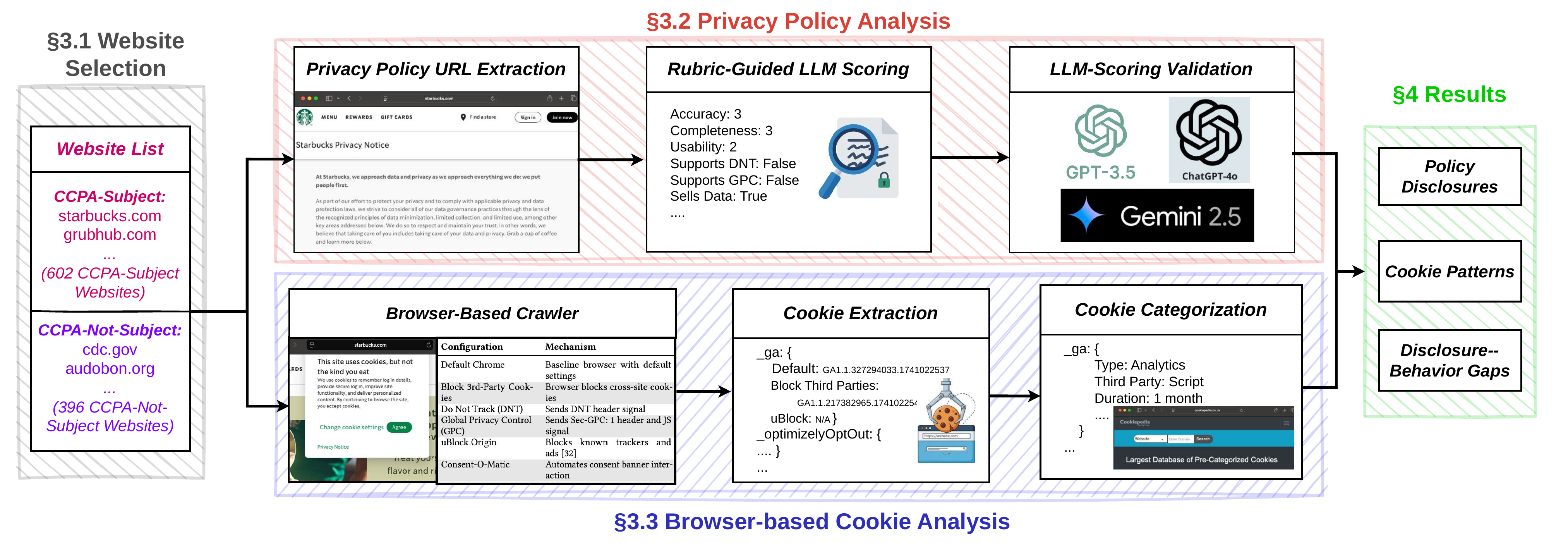}
    \caption{\textbf{\name workflow.} \name applies two parallel pipelines on websites: one that uses rubric-guided LLM scoring to analyze privacy policies, and another that uses a crawler to collect and analyze cookies.}
    \label{fig:workflow}
\end{figure*}


%% file: 3-methodology.tex
\section{Methods}
\label{sec:methodology}
\myparagraph{\name Overview}
We present \name, an automated, dual-pronged auditing framework that integrates structured LLM-based privacy policy analysis with large-scale detection of cookie writes. Figure~\ref{fig:workflow} illustrates an overview of the framework. \name takes as input a curated set of website URLs, selected to include both \subject and \notsubject websites (\S\ref{sec:website-selection}). For each site, \name executes two parallel workflows on website homepages. The first workflow, \textit{Privacy Policy Analysis}, systematically locates and extracts privacy policy links, and evaluates policies with multiple state-of-the-art LLMs, which score them based on a CCPA-guided rubric validated through expert legal review (\S\ref{sec:privacy-policy-analysis}).
To ensure robustness, we validate LLM outputs through natural-language explanations, inter-model consistency checks, and manual verification. The second workflow, \textit{Browser-based
Cookie Analysis}, employs a browser-based crawler to visit each website under six privacy configurations and collect cookies (\S\ref{sec:cookie-tracking}). Cookies are then classified based on their function and source. 
Together, the two workflows provide a novel end-to-end view of both user-facing privacy disclosures and cookie setting practices in the CCPA context.

The design of \name addresses three key challenges in privacy analysis: (1) \textit{Scalability}: traditional privacy audits require substantial manual effort to analyze policy text and site behavior. \name overcomes this through automated policy extraction, LLM-driven text analysis, and a fully automated crawling and cookie classification pipeline. (2) \textit{Legal Foundation}: prior work has rarely linked empirical analysis with the explicit requirements of privacy law. \name bridges this gap by grounding its policy evaluation rubric directly in CCPA clauses and validation  through expert legal review. (3) \textit{Realistic User Simulation}: end-users employ diverse privacy strategies (e.g., browser settings, privacy headers, consent-management extensions), making it difficult to replicate real-world browsing conditions. \name tackles this by using six distinct privacy configurations that reflect common user privacy practices, including opt-out signals~\cite{rasaii2023exploring,englehardt2015cookies}.  To ensure that our measurements reflect CCPA-relevant behaviors, all measurements were collected in a client located in California in mid-2025. 

\subsection{Website Selection}
\label{sec:website-selection}
The first step in our analysis is to identify websites relevant to our study. While measurement studies often rely on popularity-based lists, selecting domains for CCPA-focused analysis is more nuanced since the statute only applies to for-profit entities meeting specific thresholds of annual revenue and data collection scope (See \S\ref{sec:background}). However, collecting such metadata about websites is challenging at scale, as most of them do not publicly disclose corporate revenues and user demographics.

To address this challenge and ensure comparability with previous work, we base our selection on the curated data set of Tran \etal~\cite{tran2024measuring}, which provides a domain classification using structured business intelligence sources and publicly available data. We contacted the authors and obtained the dataset comprising 1,017 domains, and filtered out inactive or defunct sites, yielding a final sample of 1,002 active websites, which we refer to as the \emph{Known-Subjectivity list} (see Appendix~\ref{appendix:domain-categorization} for business classification). We re-verified and re-labeled each website's CCPA applicability from two signals used in this prior work~\cite{tran2024measuring}---company annual revenue and organizational type. We obtain revenue from the Apollo business-intelligence database~\cite{apolloio}\footnote{Tran~\etal~\cite{tran2024measuring} draw these signals from PitchBook and ZoomInfo, which are sold only through direct business-to-business sales contracts on an annual basis, which were not available to us at the time of writing. We instead subscribed to Apollo, a comparable business-intelligence provider offering self-serve access to individual researchers, and validated its revenue figures against SEC~EDGAR filings for the public companies in our sample, finding complete agreement on the statute's \$25M gross-revenue threshold.} and organizational type from public records (SEC~EDGAR~\cite{secedgar}, Wikidata~\cite{wikidata}, and GLEIF~\cite{gleif}). We exclude 4 websites whose subjectivity we cannot confidently establish, leaving \totalwebsitesbeforefiltering websites for analysis. In total, we compare 602 for-profit websites subject to the CCPA (\subject) against a group of 396 websites that are not subject (\notsubject). However, only \totalwebsites set any cookies at all upon homepage visits, so we consider only these websites for our cookies analysis.

To assess whether \name scales beyond corpora with pre-established subjectivity, we additionally construct the \emph{Popular-Website list}: a second, unlabeled set of 1,000 websites drawn directly from the Tranco ranking of popular websites~\cite{Tranco}. We keep one base domain per site (\eg \texttt{example.com}) under generic or U.S.-oriented TLDs---those most likely to be U.S.\ consumer sites relevant to the CCPA---require an active homepage, and sample 250 sites per Tranco rank band (top-1k, 1k--10k, 10k--100k, 100k--1M), excluding every domain in the Known-Subjectivity list. Unlike the Known-Subjectivity list, we do not determine whether each of these sites is subject to the CCPA; we run the full \name pipeline on them both to test scalability and to study sites that publicly claim CCPA compliance regardless of their formal subjectivity. The full protocol and results are provided in Appendix~\ref{app:tranco}.


%

\subsection{Privacy Policy Analysis}
\label{sec:privacy-policy-analysis}
A central obligation under the California Consumer Privacy Act (CCPA) is that businesses publicly disclose consumer rights such as access, deletion, and opt-out through a clear and accessible privacy policy (California Civil Code §1798.130(a)(5)(A)). In this section, we describe \name's privacy policy auditing workflow.

\subsubsection{Privacy Policy URL Extraction}
\label{sec:privacy_policy_url_extraction}
To analyze websites' alignment of privacy claims with the CCPA, it is necessary for \name to reliably locate and extract their privacy policies. However, automatically locating privacy policies across a diverse set of websites presents a unique challenge, since privacy policies are not consistently labeled or linked. Websites may use generic links such as ``Terms'', some of which may lead to irrelevant legal pages. 
Moreover, websites frequently offer multiple policy-related links, including region-specific disclosures, complicating identification. 

To resolve these challenges, we built a two-tiered automated extraction pipeline for \name using headless browsing with Selenium and XPath-based DOM querying. The pipeline first attempts a targeted XPath extraction based on a carefully curated set of high-precision privacy-related keywords that we search for in the website's homepage. To identify these keywords, we randomly sampled 50 websites from our study corpus of \totalwebsitesbeforefiltering domains and manually visited each
to identify the privacy policy URL. Using this set of labeled websites, we identified a set of two keywords---``Privacy Policy'', ``Your Privacy Rights''---that were highly successful in identifying privacy policy pages through anchor (\texttt{<a>}) tags without any false positives. 
We apply the two high-precision keywords to our full set of \totalwebsitesbeforefiltering websites. If no matching anchor (\texttt{<a>}) elements are found using this targeted approach, the pipeline proceeds to a fallback stage that scans all anchor elements for either the word ``privacy'' in their visible text or the substring ``privacy'' in their URL. This fallback ensures coverage for websites with non-standard labeling practices. Using this automated two-tiered pipeline, we were notably able to obtain privacy policy-related pages for all \totalwebsitesbeforefiltering websites in our dataset, resulting in a total of \privacypolicies combined privacy policies\footnote{We note that this number is less than \totalwebsitesbeforefiltering since many domains share privacy policies---for example, \texttt{google.com} and \texttt{youtube.com}.}, which we use for our analysis. We validated the accuracy of our method by manually visiting all privacy- and terms-related links on a separate, disjoint random sample of 50 websites from the same corpus. We only find one website (\texttt{www.underarmour.com}) with an extra page not identified with our automated pipeline, but which was focused towards international clients (not CCPA-related), which are not the focus of our study.

\subsubsection{LLM-Based Analysis of Privacy Policies}
\label{sec:llm-policy-scoring}
To audit whether websites offer actionable rights under the CCPA, \name needs to verify whether privacy policies clearly, accurately, and completely articulate user rights, while ensuring that this evaluation is robust enough to capture nuanced legal language. However, privacy policies are inherently challenging to parse, even for experts, due to their unstructured format, varied writing styles, and legalistic ambiguity~\cite{mhaidli2023researchers}. While prior NLP systems have successfully automated policy classification and practice extraction, they typically target generic privacy ontologies or specific platforms (mobile apps, VR)~\cite{cui2023poligraph,andow2020actions}. Applying these approaches to CCPA-specific compliance evaluation requires a rubric grounded in the statute's particular legal requirements.
%

To address these challenges, we design a structured, rubric-guided evaluation pipeline for \name that leverages LLMs while grounding assessments in the legal provisions of the CCPA. This approach enables \name to use LLMs as transparent, explainable evaluators rather than opaque black-box classifiers.
We begin by parsing each privacy policy into semantically distinct passages. First, we use XPath-based extraction to segment the HTML DOM into structural elements (headings, paragraphs, and list items). For policies lacking clear structural markup, we apply MiniLM sentence embeddings to cluster semantically related text spans into coherent passages. These passages are then fed into a structured LLM prompt (see Appendix~\ref{sec:full-llm-prompt} for the full prompt) that elicits a multi-dimensional assessment along with natural-language explanations for each assessment. \name's prompt asks the LLM model to evaluate six key compliance dimensions: (1) \textit{Completeness} of rights disclosure (score between 0--3), which evaluates coverage of core CCPA rights including right to access, delete, and opt-out; (2) \textit{Usability} of access mechanisms (score between 0--3), which assesses the presence of actionable mechanisms like contact forms or links; (3) \textit{Accuracy} of provisions and terminology (score between 0--3), which checks for correct use of statutory terms and definitions; (4) \textit{Disclosures} of key privacy practices (boolean), which provides binary indicators (True/False) for the presence of disclosure of data collection, sharing, purpose, retention period, and user rights (access, delete, opt out); (5) \textit{Privacy Claims} which provides ternary indicators (True/False/Unspecified) for privacy claims including the honoring of Do Not Track (DNT) and Global Privacy Control (GPC) signals, cookie setting and consent behavior, and the selling and sharing of data with third parties; and (6) \textit{Mentions} of privacy laws and practices, boolean indicators (True/False) of references to online (cookies, SDKs, analytics) and offline data collection methods, and the CCPA law itself. 

We map each of the above compliance dimensions explicitly to corresponding CCPA/CPRA statutes (Appendix~\ref{sec:legal-basis}) and validate them through expert review: an external privacy-law expert at a public-sector regulatory organization---holding a Juris Doctor with over 15 years of experience in California privacy regulation---reviewed each provision-to-dimension mapping, confirmed its correctness, and suggested minor refinements that we incorporated (see Appendix~\ref{sec:legal-basis} for details). For instance, the completeness score and disclosures of key privacy practices align with requirements under \S1798.100(a), \S1798.105, \S1798.110, \S1798.120, and \S1798.121, ensuring that policies enumerate consumer rights such as access, deletion, and opt-out of sale or sharing. 
The usability score and opt-out disclosures correspond to \S1798.130(a)\allowbreak(1)(A), which mandates that businesses provide multiple accessible methods for consumers to exercise these rights (e.g., toll-free number, webform).  
Finally, the accuracy score, privacy claims and mentions are based on \S1798.140 (definitions) and \S1798.100(b), requiring the use of correct terminology to describe rights and obligations. 
Along with each evaluation, we ask the LLM to include a natural-language rationale with direct references to the policy text. 


\subsubsection{LLM Implementation and Validation}
\label{sec:llm-validation}
Deploying LLMs for privacy policy assessments introduces a critical challenge: ensuring the reliability and interpretability of model outputs. Unlike deterministic rule-based systems, LLMs may generate variable responses due to differences in training data, architecture, and sampling behavior.  Traditional human-annotated ground truth datasets for privacy policies are scarce, making direct validation difficult. 

To resolve this challenge, we first conduct a systematic inter-model agreement analysis across three leading foundation models: GPT-3.5, GPT-4o, and Gemini 2.5. Our objective is to empirically assess whether different LLMs, when presented with the same structured prompt and policy passages, would yield consistent semantic interpretations. We treated this agreement as a practical proxy for output stability, a concept increasingly recognized in LLM evaluation literature~\cite{tang2023policygpt,rodriguez2024large}.
%
%
%
We randomly select 200 policies from our corpus and obtain independent responses from each LLM using the six-dimensional scoring schema detailed in Section~\ref{sec:llm-policy-scoring}. For categories with boolean outputs, such as \emph{Mentions} and \emph{Disclosures}, we computed Fleiss' $\kappa$ to measure inter-rater consistency across the models. The analysis yielded $\kappa = 0.74$, which falls in the
``substantial agreement'' range (0.61--0.80) on the widely used
Landis--Koch scale~\cite{landis1977measurement}. This level of
agreement is comparable to typical inter-annotator scores reported in NLP classification tasks~\cite{artstein2008survey}, and suggests that our structured prompt framework induces consistent cross-model interpretations.
For categories scored on a 0--3 scale (\eg \emph{Completeness}), we calculate pairwise Spearman rank correlations between model pairs. Correlation coefficients ranged from $\rho = 0.69$ to $\rho = 0.83$, indicating moderate to strong alignment in scoring patterns across the different LLMs. 


To further strengthen output reliability and mitigate hallucination risks, we next implement a multi-tiered validation protocol informed by the cross-model validation and manual analysis. First, all LLM outputs are required to conform to a strict JSON schema, with malformed responses automatically rejected and re-queried using reinforced prompts. Second, we manually reviewed a stratified sample of 60 outputs per model, noting hallucination rates between 2--5\% depending on policy length and model type. Common hallucinations included fabricated opt-out mechanisms or unsupported legal claims. These were addressed by iterative prompt conditioning---for example, instructing models to reply with ``Not mentioned'' when uncertain---and post-processing checks such as comparing explanations with scores and policy text: any claim not corroborated by the policy text or contradicted by the model’s own explanation is discarded. This protocol helps avoid both false positives and negatives. 
Among the three LLMs, GPT-4o provided the most consistent legal term recognition and accurate explanations, so we report results using GPT-4o with manual verification. Importantly, this selection is based on output \emph{quality}, namely closer agreement with our manual review (a random sample of 100 policies we hand-labeled as ground truth, on which GPT-4o's scores agreed with our labels 97\% of the time, versus 92\% for Gemini and 91\% for GPT-3.5) and more reliable recognition of legal terminology, and was fixed before our substantive analysis rather than chosen to fit any particular result. Moreover, the substantial cross-model agreement reported above (Fleiss $\kappa=0.74$; pairwise Spearman $\rho=0.69$--$0.83$, measured on the 200-policy sample) indicates that the three models extract largely the same signals from each policy, so the differences we report are unlikely to be specific to any single model.

To test whether residual LLM errors could affect our findings, we ran a Monte Carlo sensitivity analysis with 1{,}000 iterations, randomly flipping 5\% of boolean labels and perturbing 5\% of ordinal rubric scores by $\pm$1, matching the upper bound of our observed hallucination rate. Nine of ten originally significant results remained significant ($p<0.05$) in $\geq$99.8\% of iterations, indicating that the Subject--Not-Subject differences we report are robust to measured LLM error and stable enough for scalable auditing in practice. The only exception is the data-selling disclosure rate (33.0\% vs.\ 23.2\%, $V$=0.10), whose smaller effect size made it sensitive to label noise; we flag this in our results (\S\ref{sec:privacy-policy-results}).

\subsection{Browser-based Cookie Analysis}
\label{sec:cookie-tracking}
We next instrument \name with a browser-based crawler to audit website cookie writes, one of the most prominent methods for tracking users online~\cite{englehardt2016online}. 
Although the CCPA does not explicitly regulate cookie writes themselves,
cookies serve as the primary technical mechanism through which
third-party data collection is initiated, which \textit{can} constitute a
``sale'' or ``sharing'' of personal information under the
statute. In the landmark \textit{People v. Sephora} enforcement action (2022), the California Attorney General imposed a \$1.2 million penalty for, among other violations, failing to disclose data sales facilitated by third-party cookies and failing to honor GPC opt-out signals~\cite{sephora_ccpa_2022}. Subsequent enforcement actions have similarly targeted businesses whose cookie and opt-out practices fell short of statutory requirements~\cite{healthline_ccpa_2025,tractor_supply_cppa}. These precedents motivate our measurements of cookie deployment as a \textit{scalable and externally observable auditing signal} of data collection practices.

\subsubsection{Automated Browsing and Cookie Extraction}
\label{sec:behavioral_browsing}
We design an automated crawling infrastructure using the Chrome browser that systematically simulates real-world user visits across a diverse set of privacy configurations. While previous studies have explored website crawling and cookie analysis in detail~\cite{englehardt2016online, Degeling_2019, bollinger2022automating, matte2020cookie}, our study highlights unique challenges in simulating diverse browsing profiles and extracting structured, meaningful cookie data. In practice, users often deploy privacy-enhancing tools such as ad blockers, consent managers, or privacy headers that interact with websites and influence tracking outcomes. To simulate these behaviors, we adopt six browser configurations deliberately chosen to represent a spectrum of privacy-preserving behaviors, as shown in Figure~\ref{fig:workflow}. In the first configuration, we use the default Chrome browser with no privacy enhancements, which provides a baseline. In our second configuration, we instrument Chrome to block third-party cookies, a common user tactic to curb cross-site tracking. The third configuration involves setting the Do Not Track (DNT) header which represents a standardized, albeit often ignored~\cite{schoni2023block, libert2018automated}, signal of a user's online tracking preferences. The fourth configuration enables Global Privacy Control (GPC), a more recent privacy signal that communicates user opt-out preferences through both the \texttt{Sec-GPC: 1} HTTP header and the \texttt{navigator.global\allowbreak PrivacyControl} JavaScript API~\cite{mdn_globalPrivacyControl}. Unlike DNT, GPC carries legal weight under the CCPA, requiring subject businesses to honor the signal as a valid consumer request to opt out of data sale and sharing~\cite{ca-gpc}. This distinction makes GPC particularly relevant for our study. Finally, we incorporate two commonly-used browser extensions: \texttt{uBlock Origin} powered by EasyList~\cite{easylist} for blocking known trackers and advertisements, and \texttt{Consent-O-Matic}, which automates the rejection of cookie consent banners~\cite{nouwens2022consent}.  

To perform extensive crawling under the different privacy configurations, we use Puppeteer, a Node.js-based browser automation framework. We execute each browsing session for each privacy setting within a clean, headless Chromium instance to ensure strict session isolation. We set page load timeouts to 90 seconds or until the Document Object Model (DOM) completes loading, after which we systematically extract cookies using Puppeteer’s \texttt{page.cookies()} API. We design a structured logging system that captures both the raw cookie data and its contextual metadata within each browsing session, including the cookie name, domain, value, expiration timestamp, and setting script. 
%
We also annotate websites based on the presence of cookie consent banners, an important user-facing privacy signal made mandatory by laws like GDPR~\cite{gdpr2016regulation}. On each website homepage, we use JavaScript to automatically detect banners or consent management platforms via common identifiers and classes (e.g., ``consent'', ``onetrust-banner-sdk''). For sites with no automatic matches, we manually review the homepage to ensure all consent banners are identified. \textit{Across our dataset, only 239 (27\%) of websites displayed a consent banner}. This trend persisted among regulated entities, with just 183 (33.2\%) of \subject sites showing a banner compared to 56 (15.4\%) of \notsubject sites. Using \texttt{Consent-O-Matic}, we capture cookies under different consent scenarios when banners are present. An initial snapshot is captured immediately upon page load, representing the baseline data collection. If a consent banner is detected, we simulate user actions by both accepting and rejecting non-essential cookies using \texttt{Consent-O-Matic}, and capture follow-up snapshots of the cookie state following this interaction. 

\myparagraph{Internal-Page Crawl} To assess homepage representativeness, we additionally crawl up to five internal pages per site under all six configurations using the same analysis pipeline. Appendix~\ref{app:innerpage} provides details and results.

\subsubsection{Cookie Functionality Classification}
\label{sec:cookie_classification}
Accurate classification of cookies and understanding their purpose (\eg whether they are targeting or functional) is a critical but challenging step in analyzing tracking. Many cookies employ hashed or dynamic names and values that obscure their function. Trackers also often operate under aliased domains or content delivery networks, and individual public classification databases are often incomplete~\cite{munir2023cookiegraph, iqbal2020adgraph,dimova2021cname, papadopoulos2019cookie,amjad2021trackersift,hieu2021cv}.
To address these challenges, we synthesize datasets from multiple sources to maximize
categorization coverage. Specifically, we obtain and parse public cookie categorization datasets from Cookiepedia~\cite{cookiepedia}, the Open Cookie Database~\cite{opencookiedb}, and Cookie Cutter DB~\cite{hu2021cccc}. 
For cookies that remain unclassified after name-based matching based on these datasets, we employ domain-based categorization by matching the setting script's origin domain against known tracker databases, specifically DuckDuckGo Tracker Radar (708 categorized tracking domains)~\cite{duckduckgo_trackerradar} and Disconnect tracking protection lists (2,322 services across advertising, analytics, social, and content categories)~\cite{disconnect_tracker_lists}. 
We curate the various categories from these databases into four unique categories for our analysis: (1) \textit{Strictly Necessary} cookies (\eg \texttt{sessionID}, which maintains login sessions), (2) \textit{Functional} cookies (\eg \texttt{\_\_cf\_bm}, which Cloudflare uses to distinguish humans from bots), (3) \textit{Performance} cookies (\eg \texttt{nmstat}, which records site analytics), and (4) \textit{Targeting} cookies (\eg \texttt{FCNEC}, which is linked to Facebook and Google services for delivering targeted marketing). Any cookies for which we cannot find data across all sources are flagged as ``Unknown.''
Using our multi-source approach, \name successfully categorizes \textbf{70.6\% of all cookies} (13,502 of 19,119), leaving only 29.4\% (5,617 cookies) as ``Unknown.'' This represents a substantial improvement compared to prior academic work~\cite{munir2023cookiegraph,hu2021cccc,cahn2016empirical}. We were able to categorize at least one cookie for 877 (95.8\%) websites and at least three-quarters of cookies for 658 (72\%) websites, enabling a representative view of website behavior. 

\myparagraph{Validating Cookie Categorization}
To validate our categorization, we audited the 200 most prevalent cookies against their documented purpose in published vendor and cookie-reference sources (\eg \texttt{cookie.is} and vendor ad-tech documentation), supplemented by manual review of the most prevalent. We could establish a confident reference for 171 of them (the other 29 were too ambiguous to label). On the \textit{Targeting}-vs-non-\textit{Targeting} distinction that our findings rest on, our labels matched this reference for 94\% of the 171 (93\% weighted by prevalence), with 3 false positives and 7 false negatives. Since the errors are mostly false negatives---advertising cookies (\eg \texttt{permutive-id}, \texttt{\_\_eoi}) we conservatively mark non-Targeting, rather than consent cookies (\eg \texttt{\_tracking\_consent}) wrongly flagged---our categorization errs toward under-counting, so the tracking rates we report are a conservative lower bound.


\subsubsection{Third-Party Cookie Classification}
\label{sec:cookie-third-party-method}
We adopt a dual classification framework to distinguish first-party and third-party cookie in the modern web environment~\cite{bahrami2025cookieguard}.

\myparagraph{Domain-Based Classification}
Using the conventional definition, we classify a cookie as \textit{third-party} if its domain attribute does not match the eTLD+1 of the website. 
This classification reflects the cookie's scope of access and Same-Origin Policy restrictions, but does not reveal \textit{who} initiated the cookie write. We find only a small number of third-party cookies (4.3\%) under this definition.

\myparagraph{Script-Based Attribution}
To identify the true controller of cookie writing behavior, we adopt a \textit{script attribution} classification that determines which origin's code executed the cookie write, regardless of the cookie's domain. Our approach builds on established call-stack interception techniques used for cookie attribution in prior work~\cite{englehardt2016online,chen2021cookie,bahrami2025cookieguard}; we apply these techniques across our six configurations to measure how third-party script behavior changes in response to user privacy practices. Specifically, we instrument two interception methods: first, we redefine the native \texttt{document.cookie} setter to capture every JavaScript-initiated cookie write along with its call stack, including the \textit{source URL} of each function in the stack; second, we monitor all HTTP responses for \texttt{Set-Cookie} headers to capture server-side cookies, including those marked \texttt{HttpOnly}.

A cookie is then classified as \textit{third-party by attribution} if the source URL of the script that initiated the write belongs to a different eTLD+1 than the visited website, even when the cookie itself carries a first-party domain. For example, if a script loaded from \texttt{google\allowbreak-analytics.com} writes a cookie scoped to \texttt{.example.com} via \texttt{document.cookie}, our call-stack instrumentation technique records \texttt{google\allowbreak-analytics.com} as the initiating origin, and we classify this cookie as third-party because the external script controls the cookie lifecycle.
This distinction is critical because third-party scripts routinely set nominally first-party cookies to evade browser restrictions on third-party cookie access while maintaining cross-site tracking capabilities~\cite{chen2021cookie, munir2023cookiegraph, demir2022towards, bahrami2025cookieguard}. We note that our method attributes cookies based on the \textit{URL from which the browser loaded the script}; if a website self-hosts or proxies a third-party script through its own domain (\eg via CNAME cloaking~\cite{dimova2021cname}), the script would appear as first-party in our classification. This is a known limitation shared with prior call-stack-based attribution approaches~\cite{chen2021cookie, munir2023cookiegraph}, and means our third-party results represent a conservative lower bound. An independent re-parse of the raw call stacks reproduced our attribution for 99.97\% (7{,}358/7{,}360) of script-set cookies with parseable stacks.

\subsection{Statistical Analysis}
\label{sec:statistical-analysis}
We evaluate the significance of differences between \subject and \notsubject websites using non-parametric and categorical tests. For ordinal and count-based metrics such as rubric scores (rated 0--3) and cookie counts, we use the Mann\allowbreak–Whitney $U$ test, which does not assume normality and fits our distributions skewed towards higher values for \subject websites. For binary and categorical variables, including disclosures, behavioral claims, and third-party script attribution, we use Pearson's $\chi^2$ test of independence, and apply Fisher’s exact test when expected counts are below five. We report two-tailed $p$-values and set the significance level to $\alpha=0.05$. For interpretability, we include effect sizes: the rank-biserial correlation ($r$) for Mann--Whitney tests and Cramér's~$V$ for $\chi^2$ tests. Values of 0.1, 0.3, and 0.5 correspond to small, medium, and large effects respectively.

Because we run many comparisons, we treat all 25 tests as a single family and report multiple-testing--adjusted $p$-values two ways: the Benjamini--Hochberg procedure~\cite{benjamini1995controlling}, controlling the false-discovery rate (FDR), and---following prior CCPA measurement work~\cite{o2021clear}---the stricter Holm--Bonferroni step-down procedure~\cite{holm1979simple}, controlling the family-wise error rate (FWER). Holm--Bonferroni is a uniformly more powerful form of the classic Bonferroni correction, and its family-wise error control holds under arbitrary dependence among the tests~\cite{holm1979simple}; we base our significance decisions on the more conservative Holm-adjusted $p$-values, with Benjamini--Hochberg as a complementary view. Appendix~\ref{sec:multiple-hypothesis} reports the raw and adjusted $p$-values for all 25 tests. Under the stricter Holm criterion, our main takeaways in \S\ref{sec:results} remain unchanged.
%
%
To confirm these disclosure differences reflect CCPA applicability rather than firm size or ad-revenue reliance, we additionally adjust for both via per-disclosure logistic regressions and a size-matched comparison against large CCPA-exempt organizations with full specifications and results in \S\ref{sec:privacy-policy-results} and Appendix~\ref{app:size-control}.

%% file: 4-results.tex
\section{Results}
\label{sec:results}
We present \name's findings in three parts: privacy policy auditing (\S\ref{sec:privacy-policy-results}), cookie-write auditing (\S\ref{sec:tracking-results}), and a joint policy--behavior analysis that links disclosures to observed tracking (\S\ref{sec:policy-behavior-results}). We additionally validate that these signals generalize to an unlabeled 1,000-site corpus in a scaling case study (Appendix~\ref{app:tranco}).

\subsection{Privacy Policy Findings}
\label{sec:privacy-policy-results}
\input{figures/rubric-scores-subjectivity}

\input{tables/policy-results-overall}
We analyze the six key compliance dimensions across \subject ($N=567$) and \notsubject ($N=366$) policies in the \privacypolicies privacy policies identified by \name.
First, we observe that \subject privacy policies are significantly more complete (Mann--Whitney $U$ test, $p<0.001$, $r=0.27$), usable ($p<0.001$, $r=0.31$), and accurate ($p<0.001$, $r=0.25$) than \notsubject policies, as shown in Figure~\ref{fig:rubric_scores_subjectivity}.
The median score for \subject policies is 3 across all three dimensions, compared to 2 for \notsubject policies, indicating that CCPA-subject websites tend to provide more detailed, actionable, and legally precise disclosures.
\name's LLM-audit noted specific examples of this: [Completeness] ``The policy clearly lists consumer rights under CCPA, including access, deletion, and opt-out of sale/share, with detailed instructions on how to exercise these rights.'', [Usability] ``The policy provides multiple clear methods to exercise rights, including web links and a phone number, with detailed guidance on verification and timelines.'', and [Accuracy] ``The policy uses correct legal terminology and aligns with CCPA/CPRA language, including references to GPC and opt-out rights''.  In contrast, \notsubject policies scored lower, especially in usability and completeness. For example, ``The policy mentions opt-out options for targeted advertising and data removal but lacks detailed instructions on exercising rights like access or deletion.'' We noted that $\sim$31\% of \notsubject policies missed details regarding specific CCPA rights such as the right to access and delete data. Over 40\% of \notsubject policies lacked detailed opt-out instructions, and these policies generally scored lower on usability.

While we observe CCPA-subject sites exhibiting better disclosure practices, some of this difference may stem from these sites being larger and deriving more revenue from data sharing. Controlling for company revenue and third-party tracking intensity, however, CCPA subjectivity remains a strong predictor of the opt-out, access, and deletion disclosures (all $p<0.001$), and the gap persists among large organizations above the statute's \$25M threshold (opt-out 77\% vs.\ 46\% for large but exempt nonprofits and government agencies); we detail this in Appendix~\ref{app:size-control}. The gap thus reflects CCPA \emph{applicability} rather than firm size or ad-revenue reliance alone, though as an observational comparison it cannot establish strict causation. Consistent with this, 67.3\% of \subject privacy policies mention the CCPA (Table~\ref{tab:policy-results-overall}), often with sections addressing its specific clauses, and 43.1\% of \notsubject sites also mention it---a notable spillover effect---suggesting the growth of transparency is attributable at least in part to the CCPA, in line with prior work~\cite{tran2024measuring} showing more opt-out links when sites are accessed from California.

\takeaway{\subject websites provide significantly stronger privacy disclosures than \notsubject websites across completeness, usability, and accuracy.}

Table~\ref{tab:policy-results-overall} shows that many core disclosures are common across both groups, but \subject policies are stronger on CCPA-specific rights. More than 85\% of both \subject and \notsubject policies disclose data collection and sharing practices. However, the specificity of these statements varies substantially, with some specifying exactly what data is collected and shared (\eg ``The information we collect [includes]...interactions, device information, [and] location information'') and others providing generic information (\eg ``We collect several different types of information for various purposes to provide and improve our Service to you''). Importantly, we find that \subject policies are significantly more likely to disclose the right to access data (85.4\% vs.\ 68.3\%, $\chi^2$, $p<0.001$, $V=0.20$), the right to delete data (85.9\% vs.\ 69.4\%, $p<0.001$, $V=0.20$), and opt-out instructions (77.1\% vs.\ 57.1\%, $p<0.001$, $V=0.21$). We note that the percentage of \subject websites offering opt-out information and links has slightly increased from previous studies such as Tran~\etal~\cite{tran2024measuring}, which reported only $\sim$70\% of \subject websites included an opt-out link. \subject websites are also significantly more likely to disclose data retention details than \notsubject websites (64.6\% vs.\ 49.2\%, $\chi^2$, $p<0.001$, $V=0.15$), though retention remains one of the weaker disclosure categories overall. Finally, online data collection practices are widely disclosed across both groups (93.2\% overall), while offline data collection is much less commonly disclosed (26.0\% overall).

When it comes to behavioral claims, we observe that both \subject and \notsubject privacy policies do not specify detailed privacy-related practices (Table~\ref{tab:policy-results-overall}). For example, while most privacy policies (86.0\%) mention cookies, less than 2\% provide detailed information regarding cookies and consent behavior. We show later in \S\ref{sec:tracking-results} that both \subject and \notsubject websites set a significant quantity of Targeting cookies despite user consent action, emphasizing the need to disclose such tracking behaviors. Similarly, more than 80\% of both \subject and \notsubject policies disclose sharing personal information with third parties, but explicit data-selling disclosures are less common (33.0\% vs.\ 23.2\%, $\chi^2$, $p=0.002$, $V=0.10$), though we note that this finding is sensitive to residual LLM labeling error (\S\ref{sec:llm-validation}). Finally, a significantly larger portion (28.7\%) of \subject websites claim to honor the Global Privacy Control (GPC) signal compared to \notsubject websites (13.1\%) ($\chi^2$, $p<0.001$, $V=0.18$).
Interestingly, a few websites specifically mention not honoring this signal mandated by the CCPA, \eg ``We do not currently support the Do Not Track browser or any Global Opt Out/Global Privacy Control option''.  Similarly, a significant portion of privacy policies (31.6\%) in both cases specifically mention not honoring the DNT signal for storing cookies. 

All of our main policy findings survive multiple-testing adjustment under the stricter Holm procedure (Holm-adjusted $p$ from ${<}10^{-17}$ to $0.016$, the latter for the data-selling disclosure; Appendix~\ref{sec:multiple-hypothesis}). The comparisons that do not survive are ones we do not treat as findings, like broad disclosures that are common across both groups (\eg general data collection, data sharing, and purpose of collection) and several secondary cookie tests (\eg the overall third-party rate).

\takeaway{\subject policies more often disclose access, deletion, and opt-out rights, while broad disclosures remain common across both groups. However, very few policies describe specific privacy practices in detail.}

\subsection{Cookies Analysis}
\label{sec:tracking-results}
\input{figures/new_figures/split_violin_subject_vs_non}
We next analyze cookies across the cookie-setting \totalwebsites websites.

\myparagraph{Data Collection by CCPA Subjectivity} Figure~\ref{fig:split-violin-subj-nonsubj} shows that \subject and \notsubject websites exhibit broadly similar cookie-writing behavior at first visit.  In the Default configuration, both groups set a median of 3 Targeting cookies and 3 Performance cookies per website. Mann--Whitney tests show no significant difference in Targeting cookies between \subject and \notsubject websites ($p=0.56$, $r=0.02$), revealing that both sets of websites set advertising and tracking cookies such as \texttt{fbp} (Meta/Facebook) and \texttt{MUID} (Microsoft/Bing Identifiers). 
Across cookie categories, all effect sizes are small ($r<0.3$), suggesting that CCPA subjectivity is not strongly associated with reduced front-end cookie collection.
Unfortunately, as noted in our privacy policy results (\S\ref{sec:privacy-policy-results}), detailed explanations of cookie use are also often absent from privacy policies, reducing overall transparency. We note that any observed differences in cookie deployment between \subject and \notsubject could be influenced by business characteristics: \subject websites tend to be larger commercial entities with more complex advertising relationships and greater reliance on third-party analytics.

Figure~\ref{fig:split-violin-subj-nonsubj} and Table~\ref{tab:comprehensive-third-party-analysis} also show the distribution of cookies set across the various privacy configurations excluding \texttt{Consent-O-Matic}, which we evaluate separately later. Privacy configurations reduce cookies to different degrees, but none eliminate tracking. Blocking third-party cookies reduces total cookies by 33.4\%, while \textit{DNT reduces total cookies by 37.1\% and Targeting cookies by 38.1\%}. This shows that the DNT signal is still respected by websites for setting cookies, contrary to findings from previous work~\cite{libert2018automated}.

\textit{The GPC signal demonstrates even stronger effectiveness, achieving an overall reduction of 51\% in cookies, with Targeting cookies decreasing by 55.8\% and Performance cookies by 33.4\%.} However, GPC's effectiveness remains incomplete: even among \subject websites, which are legally required under CCPA to honor GPC signals, about 45\% of Targeting cookies are still set when GPC is enabled.

%
\textit{The most significant effect is observed in the uBlock configuration}, where the EasyList blocklist~\cite{easylist} results in a decrease of 87\% in Targeting cookies and 82\% in Performance cookies. Third-party blocking lists are extremely successful in blocking non-functional cookies, preventing targeting but maintaining usability.

\takeaway{Cookie writes are similar across \subject and \notsubject websites, and privacy signals such as GPC reduce but do not eliminate Targeting cookies.}

\input{tables/third_party_comprehensive}
\myparagraph{Third-Party Cookies} Out of the 19,119 total cookies we analyze, we find that \textit{58.5\% of cookies (11,176) are purely first-party, whereas 41.5\% (7,943) involve third-party writes}, identified through either domain-based or script-based attribution (Table~\ref{tab:comprehensive-third-party-analysis}). The substantial presence of third-party cookies shows that third-parties still collect information about users from a large number of websites: \textit{665 websites (72.8\%) set at least one third-party cookie}.
In total, 40.4\% (7,729 cookies) are discovered through our script attribution technique, compared to only 4.3\% of cookies (823) discovered through domain-based classification, with an overlap of 3.2\% of cookies. This discrepancy shows that most third-party cookie activity occurs through scripts that write cookies scoped to first-party domains, a pattern that can evade conventional browser-level third-party cookie blocking~\cite{chen2021cookie,munir2023cookiegraph,demir2022towards,bahrami2025cookieguard}.

\input{tables/top-cookie-scripts}
\myparagraph{Concentration Among Dominant Platforms} Third-party script-set Targeting cookies are \textit{highly concentrated} among a small number of major technology platforms.  Table~\ref{tab:top-cookie-scripts} reveals that just 10 third-party scripts account for 58.4\% of all script-set targeting cookies. Google's advertising and tag-management infrastructure alone accounts for 576 Targeting cookies across 386 websites (42.1\% of our sample), followed by major platforms such as Microsoft (275 cookies), Meta (197 cookies), and Adobe (181 cookies). 
This concentration suggests that a handful of intermediaries control the front-end data collection infrastructure across our study websites.

\myparagraph{Third-Party Attribution by CCPA Applicability.}  Within this concentrated ecosystem, \subject and \notsubject websites exhibit comparable overall third-party script-set cookie rates (40.0\% vs.\ 41.1\%, $\chi^2$, n.s.), as shown in Table~\ref{tab:comprehensive-third-party-analysis}.  The difference is most pronounced for Targeting cookies: \textit{51.0\% of Targeting cookies on \subject websites are set by third-party scripts}, compared to 45.0\% on \notsubject websites ($p<0.001$, $V=0.06$). In contrast, \notsubject websites show higher third-party rates for Performance and Functional cookies. Because third-party Targeting cookies are a common initial mechanism for user data collection and cross-site tracking, their prevalence among \subject websites provides an important signal for auditing.

\myparagraph{Privacy Configuration Effects on Third-Party Tracking}
Privacy-enhancing configurations demonstrate varying effectiveness in mitigating third-party cookies, as shown in Table~\ref{tab:comprehensive-third-party-analysis}. The uBlock configuration achieves the most substantial reduction in third-party cookies, reducing by 82\% for \subject websites and 89\% for \notsubject websites. The GPC signal also demonstrates meaningful effectiveness, reducing third-party cookies by 37\% for \subject websites and 31\% for \notsubject websites. However, GPC's effectiveness remains incomplete: \textit{despite GPC's legal mandate under CCPA for \subject businesses to honor opt-out requests, GPC reduces third-party Targeting cookies by only 41\% for \subject websites}. Our study reveals that the GPC disproportionately results in reduced true first-party cookies (66\% reduction) while retaining more cookies set by third-party scripts (only 40.3\% reduction). This persistence matters because third-party scripts executing on a visited site can read first-party-scoped cookies locally and transmit identifiers to their own servers, enabling cross-site correlation through identifier smuggling~\cite{bahrami2025cookieguard,randall2022measuring}.

\takeaway{Third-party tracking on \subject websites is widespread, script-driven, and concentrated.}

\subsection{Auditing Disclosures vs. Tracking}
\label{sec:policy-behavior-results}
Finally, we link \name's policy audits with per-website cookie measurements to identify audit signals that combine what websites disclose with what users can observe at first visit. 

\myparagraph{GPC disclosures} Websites that claim to honor GPC often continue setting Targeting cookies when GPC is enabled. Among \subject policies, 28.7\% (163 sites) claim to honor GPC. Of these, 151 have matched cookie data, and 110 set Targeting cookies in both Default and GPC configurations. Among these 110 sites, 52.7\% (58 sites) show less than a 20\% reduction in Targeting cookies under GPC, and 39.1\% (43 sites) show no reduction at all. The gap is even more pronounced for third-party Targeting cookies: among sites that set third-party Targeting cookies in both configurations, 66.2\% show less than a 20\% reduction, and 47.5\% show no reduction. These disclosure--behavior gaps provide a strong audit signal because they resemble issues raised in prior enforcement actions such as \textit{People v.\ Sephora}, which involved alleged sale of personal information through online tracking technologies, including cookies, and failure to process GPC opt-out requests~\cite{sephora_ccpa_2022}. The same gap appears among \notsubject websites: of the 46 that claim to honor GPC and set cookies, 67\% (31) still set at least one third-party Targeting cookie at initial load, and 61\% do so even when GPC is sent.

\myparagraph{Data selling disclosures} 204 (36.0\%) \subject policies explicitly state they do not sell personal data. Of the 187 such sites that set cookies, 50.8\% (95 sites) nonetheless deploy third-party targeting cookies at initial load, and 23.5\% (44 sites) set five or more such cookies. While cookies alone do not constitute a ``sale'' under the CCPA (\S\ref{sec:cookie-tracking}), third-party targeting cookies are the primary technical mechanism through which cross-site data sharing is initiated~\cite{englehardt2016online}, making their co-occurrence with no-sale disclosures an important signal for further manual review.

\myparagraph{Required disclosures} The CCPA requires a policy to disclose the categories of personal information a business collects (\S1798.130\allowbreak(a)\allowbreak(5)\allowbreak(B)\allowbreak(i)) and those it sells or shares, or a statement that it does neither (\S1798.130\allowbreak(a)\allowbreak(5)\allowbreak(C)\allowbreak(i)). Combining both lenses lets us check these mandates against observed behavior: of the 269 \subject websites that set third-party Targeting cookies at initial load, 26 (9.7\%) do not disclose any data collection, and 22 (8.2\%) disclose neither the sale nor the sharing of personal information (29 under a stricter reading). Because setting third-party Targeting cookies is precisely the collection and sharing these provisions ask businesses to disclose, the co-occurrence of tracking with absent required disclosures is an important signal for
further manual review.


\myparagraph{Policy quality} High-quality privacy policies do not necessarily correspond to lower cookie-based tracking. Even among the 394 \subject websites in our dataset with high-quality privacy policies (completeness and usability scores $\geq$2), 43.7\% (172 sites) set five or more Targeting cookies on first visit. The median Targeting cookie count for these high-quality-policy sites is identical to the population median, indicating that stronger disclosures are not associated with reduced cookie-based tracking.


\takeaway{\subject websites place third-party tracking cookies even when policies claim to honor privacy signals, state they do not sell data, or provide high-quality disclosures.}

\input{tables/thirdpart_cookie_attribution}\input{tables/cookie-setter-breakdown}

%% file: figures/rubric-scores-subjectivity.tex
\begin{figure}[t]
    \centering
    \includegraphics[width=0.9\linewidth]{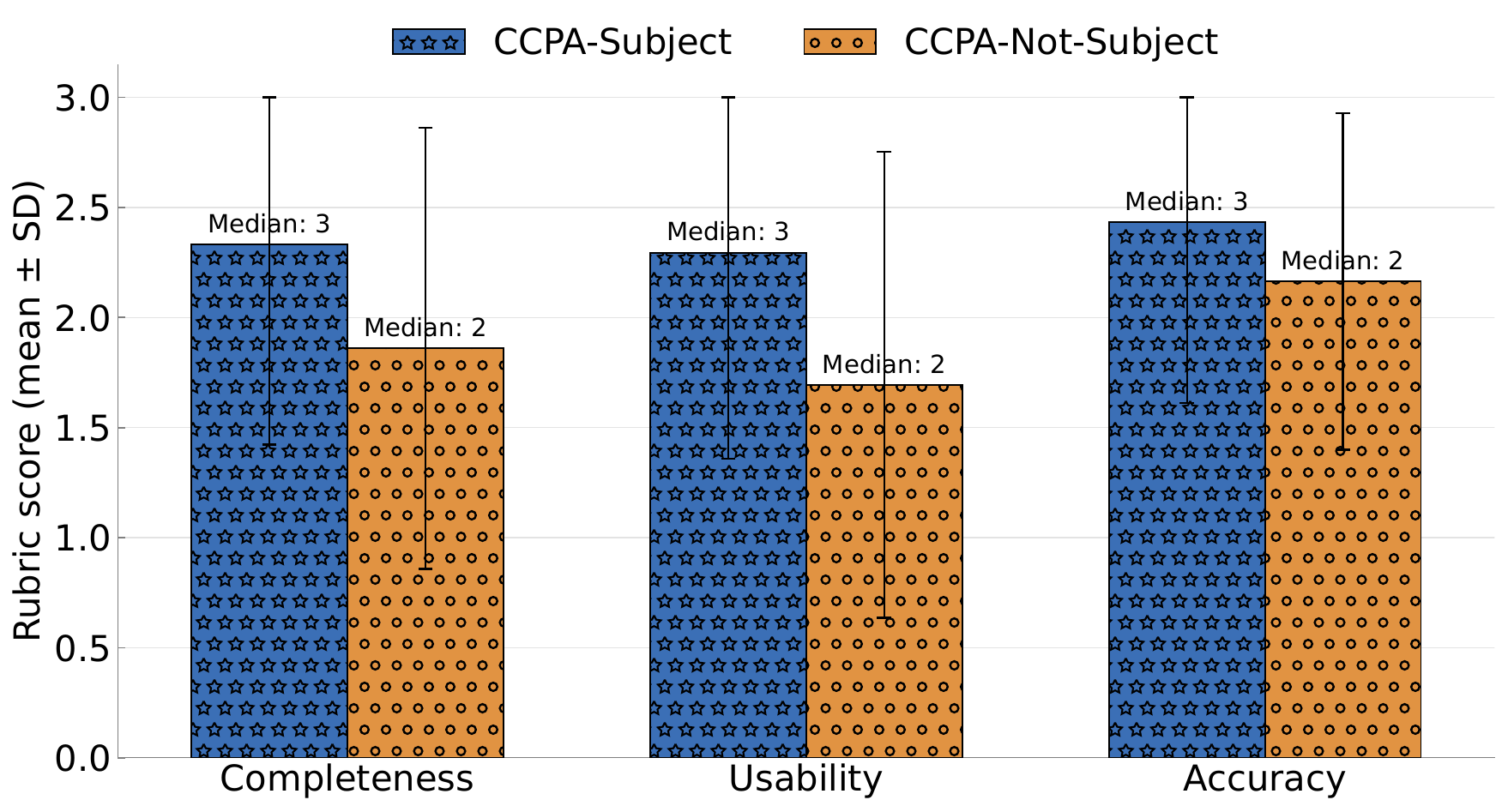}
    \caption{Rubric scores (Completeness, Usability, Accuracy) of privacy policies by CCPA subjectivity. 
    Bars show group means with error bars (standard deviation).}
    \label{fig:rubric_scores_subjectivity}
\end{figure}

%% file: tables/policy-results-overall.tex
\begin{table*}[!t]
\centering

\caption{Disclosure coverage, behavioral claims, and privacy-practice mentions in privacy policies. “True” percentages are shown for disclosures and mentions; behavioral claims report True (T), False (F), and Unspecified (U) rates.}
\label{tab:policy-results-overall}
\scriptsize
\begin{tabular}{p{2.5cm}|p{1.3cm}|p{1.3cm}|p{10.8cm}}
\toprule
\textbf{Metric / Claim} & \textbf{\texttt{Subject}} & \textbf{\texttt{Not-Subject}} & \textbf{Example Policy Excerpt (True cases)} \\
\midrule
\textbf{Disclosure Coverage} & \textbf{$n=567$} & \textbf{$n=366$} &  \\
\hline
Data Collection
                                    & \gradientcell{85.9}{1}{100}{red}{green}{40}\%
                                    & \gradientcell{89.9}{1}{100}{red}{green}{40}\%
                                    & ``The information we collect [includes]...interactions, device information, [and] location information'' \\
\cellcolor{gray!20}Data Sharing
                                    & \gradientcell{86.4}{1}{100}{red}{green}{40}\%
                                    & \gradientcell{83.3}{1}{100}{red}{green}{40}\%
                                    &\cellcolor{gray!20}``We may share your information with our service providers… advertisers...[and] affiliates'' \\
Data Collection Opt Out
                                    & \gradientcell{77.1}{1}{100}{red}{green}{40}\%
                                    & \gradientcell{57.1}{1}{100}{red}{green}{40}\%
                                    & ``[C]ontrol whether [this website] shares your personal information...by using the 'Data sharing...' option'' \\
\cellcolor{gray!20}Purpose of Collection
                                    & \gradientcell{87.3}{1}{100}{red}{green}{40}\%
                                    & \gradientcell{91.0}{1}{100}{red}{green}{40}\%
                                    &\cellcolor{gray!20}``We use the information we collect to provide and operate [feature]...improve and personalize [feature]...foster safety and security...[and] measure and analyze'' \\
Data Retention Period
                                    & \gradientcell{64.6}{1}{100}{red}{green}{40}\%
                                    & \gradientcell{49.2}{1}{100}{red}{green}{40}\%
                                    & ``We keep different types of information for different periods...cookies up to 13 months...'' \\
\cellcolor{gray!20}Right to Access
                                    & \gradientcell{85.4}{1}{100}{red}{green}{40}\%
                                    & \gradientcell{68.3}{1}{100}{red}{green}{40}\%
                                    &\cellcolor{gray!20}``You can access, correct, or modify the information...You can download a copy of your information'' \\
Right to Delete
                                    & \gradientcell{85.9}{1}{100}{red}{green}{40}\%
                                    & \gradientcell{69.4}{1}{100}{red}{green}{40}\%
                                    & ``If you follow the instructions... your account will be deactivated and your data will be queued for deletion.'' \\
\midrule
\textbf{Mentions} & \textbf{$n=567$} & \textbf{$n=366$} & \\
\hline
\cellcolor{gray!20}Cookies
                                    & \gradientcell{84.1}{1}{100}{red}{green}{40}\%
                                    & \gradientcell{89.1}{1}{100}{red}{green}{40}\%
                                    &\cellcolor{gray!20}``We use cookies, pixels and other Tracking Technologies to collect information about you'' \\
CCPA
                                    & \gradientcell{67.3}{1}{100}{red}{green}{40}\%
                                    & \gradientcell{43.1}{1}{100}{red}{green}{40}\%
                                    & ``California residents can submit requests to opt out of the sale of personal information under the...(CCPA). '' \\
\cellcolor{gray!20}Online Data Collection
                                    & \gradientcell{92.1}{1}{100}{red}{green}{40}\%
                                    & \gradientcell{94.5}{1}{100}{red}{green}{40}\%
                                    &\cellcolor{gray!20}``We may send...cookies...We may also use other similar technologies such as tracking pixels, tags...'' \\
Offline Data Collection
                                    & \gradientcell{35.8}{1}{100}{red}{green}{40}\%
                                    & \gradientcell{10.7}{1}{100}{red}{green}{40}\%
                                    & ``Personal information may be collected ...when you visit our stores... or deal with customer service'' \\
\midrule
\textbf{Behavioral Claims} & \textbf{$n=567$} & \textbf{$n=366$} & \\
\hline
\cellcolor{gray!20}Deletes Cookie After Rejecting Consent
                                    &\begin{tabular}[t]{@{}p{1.3cm}}
                                        \gradientcellwtext{0.2}{0}{100}{red}{green}{40}{T: 0.2\%}\\
                                        \gradientcellwtext{0.0}{0}{100}{red}{green}{40}{F: 0.0\%}\\
                                        \gradientcellwtext{99.8}{0}{100}{red}{green}{40}{U: 99.8\%}
                                    \end{tabular}
                                    &\begin{tabular}[t]{@{}p{1.3cm}}
                                        \gradientcellwtext{0.3}{0}{100}{red}{green}{40}{T: 0.3\%}\\
                                        \gradientcellwtext{0.0}{0}{100}{red}{green}{40}{F: 0.0\%}\\
                                        \gradientcellwtext{99.7}{0}{100}{red}{green}{40}{U: 99.7\%}
                                    \end{tabular}
                                    &\cellcolor{gray!20}``[C]lick here to...control, disable, or delete [cookies].'' \\
Sets Cookies After Rejecting Consent
                                    &\begin{tabular}[t]{@{}p{1.3cm}}
                                        \gradientcellwtext{0.0}{0}{100}{red}{green}{40}{T: 0.0\%}\\
                                        \gradientcellwtext{0.2}{0}{100}{red}{green}{40}{F: 0.2\%}\\
                                        \gradientcellwtext{99.8}{0}{100}{red}{green}{40}{U: 99.8\%}
                                    \end{tabular}
                                    &\begin{tabular}[t]{@{}p{1.3cm}}
                                        \gradientcellwtext{0.0}{0}{100}{red}{green}{40}{T: 0.0\%}\\
                                        \gradientcellwtext{0.3}{0}{100}{red}{green}{40}{F: 0.3\%}\\
                                        \gradientcellwtext{99.7}{0}{100}{red}{green}{40}{U: 99.7\%}
                                    \end{tabular}
                                    & No true cases. \\
\cellcolor{gray!20}Sets Cookies Before Consent
                                    &\begin{tabular}[t]{@{}p{1.3cm}}
                                        \gradientcellwtext{0.7}{0}{100}{red}{green}{40}{T: 0.7\%}\\
                                        \gradientcellwtext{1.4}{0}{100}{red}{green}{40}{F: 1.4\%}\\
                                        \gradientcellwtext{97.9}{0}{100}{red}{green}{40}{U: 97.9\%}
                                    \end{tabular}
                                    &\begin{tabular}[t]{@{}p{1.3cm}}
                                        \gradientcellwtext{1.4}{0}{100}{red}{green}{40}{T: 1.4\%}\\
                                        \gradientcellwtext{3.0}{0}{100}{red}{green}{40}{F: 3.0\%}\\
                                        \gradientcellwtext{95.6}{0}{100}{red}{green}{40}{U: 95.6\%}
                                    \end{tabular}
                                    &\cellcolor{gray!20}``If you continue using [website] we will assume that you are happy to receive cookies.'' \\
Tracking Only After Consent
                                    &\begin{tabular}[t]{@{}p{1.3cm}}
                                        \gradientcellwtext{1.2}{0}{100}{red}{green}{40}{T: 1.2\%}\\
                                        \gradientcellwtext{0.4}{0}{100}{red}{green}{40}{F: 0.4\%}\\
                                        \gradientcellwtext{98.4}{0}{100}{red}{green}{40}{U: 98.4\%}
                                    \end{tabular}
                                    &\begin{tabular}[t]{@{}p{1.3cm}}
                                        \gradientcellwtext{1.6}{0}{100}{red}{green}{40}{T: 1.6\%}\\
                                        \gradientcellwtext{0.0}{0}{100}{red}{green}{40}{F: 0.0\%}\\
                                        \gradientcellwtext{98.4}{0}{100}{red}{green}{40}{U: 98.4\%}
                                    \end{tabular}
                                    & ``By default, you are opted out of all cookie categories except strictly necessary cookies.'' \\
\cellcolor{gray!20}Sells Data
                                    &\begin{tabular}[t]{@{}p{1.3cm}}
                                        \gradientcellwtext{33.0}{0}{100}{red}{green}{40}{T: 33.0\%}\\
                                        \gradientcellwtext{36.0}{0}{100}{red}{green}{40}{F: 36.0\%}\\
                                        \gradientcellwtext{31.0}{0}{100}{red}{green}{40}{U: 31.0\%}
                                    \end{tabular}
                                    &\begin{tabular}[t]{@{}p{1.3cm}}
                                        \gradientcellwtext{23.2}{0}{100}{red}{green}{40}{T: 23.2\%}\\
                                        \gradientcellwtext{42.1}{0}{100}{red}{green}{40}{F: 42.1\%}\\
                                        \gradientcellwtext{34.7}{0}{100}{red}{green}{40}{U: 34.7\%}
                                    \end{tabular}
                                    &\cellcolor{gray!20}``We may disclose certain personal information in exchange for services, insights, or other valuable consideration.'' \\
Shares Data with Third Parties
                                    &\begin{tabular}[t]{@{}p{1.3cm}}
                                        \gradientcellwtext{82.7}{0}{100}{red}{green}{40}{T: 82.7\%}\\
                                        \gradientcellwtext{0.2}{0}{100}{red}{green}{40}{F: 0.2\%}\\
                                        \gradientcellwtext{17.1}{0}{100}{red}{green}{40}{U: 17.1\%}
                                    \end{tabular}
                                    &\begin{tabular}[t]{@{}p{1.3cm}}
                                        \gradientcellwtext{79.8}{0}{100}{red}{green}{40}{T: 79.8\%}\\
                                        \gradientcellwtext{4.4}{0}{100}{red}{green}{40}{F: 4.4\%}\\
                                        \gradientcellwtext{15.8}{0}{100}{red}{green}{40}{U: 15.8\%}
                                    \end{tabular}
                                    & ``We may share your personal information with...service providers'' \\
\cellcolor{gray!20}Honors GPC
                                    &\begin{tabular}[t]{@{}p{1.3cm}}
                                        \gradientcellwtext{28.7}{0}{100}{red}{green}{40}{T: 28.7\%}\\
                                        \gradientcellwtext{0.2}{0}{100}{red}{green}{40}{F: 0.2\%}\\
                                        \gradientcellwtext{71.1}{0}{100}{red}{green}{40}{U: 71.1\%}
                                    \end{tabular}
                                    &\begin{tabular}[t]{@{}p{1.3cm}}
                                        \gradientcellwtext{13.1}{0}{100}{red}{green}{40}{T: 13.1\%}\\
                                        \gradientcellwtext{0.8}{0}{100}{red}{green}{40}{F: 0.8\%}\\
                                        \gradientcellwtext{86.1}{0}{100}{red}{green}{40}{U: 86.1\%}
                                    \end{tabular}
                                    &\cellcolor{gray!20}``You may use the Global Privacy Control (GPC)...If GitHub detects the GPC signal from your device, GitHub will not share your data.'' \\
Honors DNT
                                    &\begin{tabular}[t]{@{}p{1.3cm}}
                                        \gradientcellwtext{2.1}{0}{100}{red}{green}{40}{T: 2.1\%}\\
                                        \gradientcellwtext{33.5}{0}{100}{red}{green}{40}{F: 33.5\%}\\
                                        \gradientcellwtext{64.4}{0}{100}{red}{green}{40}{U: 64.4\%}
                                    \end{tabular}
                                    &\begin{tabular}[t]{@{}p{1.3cm}}
                                        \gradientcellwtext{1.9}{0}{100}{red}{green}{40}{T: 1.9\%}\\
                                        \gradientcellwtext{27.6}{0}{100}{red}{green}{40}{F: 27.6\%}\\
                                        \gradientcellwtext{70.5}{0}{100}{red}{green}{40}{U: 70.5\%}
                                    \end{tabular}
                                    & ``If your browser sends a Do Not Track (DNT) signal, GitHub will not set non-essential cookies and will not load third party resources'' \\
\bottomrule
\end{tabular}%
\end{table*}

%% file: figures/new_figures/split_violin_subject_vs_non.tex
\begin{figure*}[!t]
    \centering
    \includegraphics[width=\linewidth]{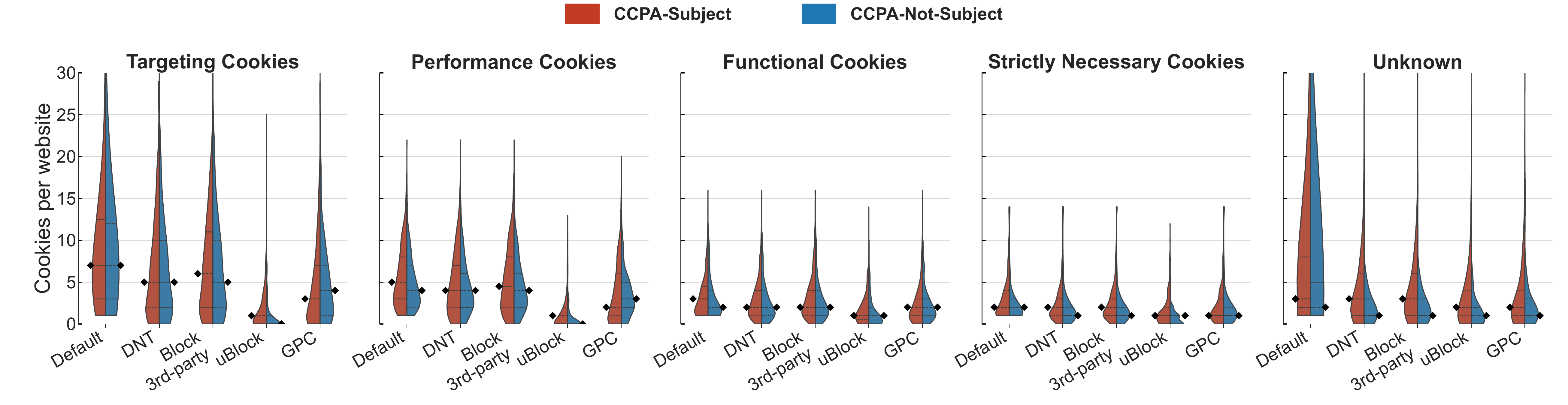}
    \caption{
        Cookie counts per website separated by CCPA subjectivity, cookie types, and browsing configurations.
    }
    \label{fig:split-violin-subj-nonsubj}
\end{figure*}

%% file: tables/third_party_comprehensive.tex
\begin{table}[!t]
\caption{Third-party cookies by CCPA subjectivity. Default state shows cookie counts and third-party attribution percentages. Privacy
configurations show percentage reduction in ``total cookies (third-party cookies)''.}
\label{tab:comprehensive-third-party-analysis}
\centering
\scriptsize
\setlength{\tabcolsep}{1pt}

\resizebox{\columnwidth}{!}{
\begin{tabular}{@{}lrrrcccc@{}}
\toprule
\textbf{Cookie Type} & \textbf{N} & \textbf{Script} & \textbf{Domain} & \multicolumn{4}{c}{\textbf{↓\% Total (↓\% 3P)}} \\
\cmidrule(lr){5-8}
 & & \textbf{3P\%} & \textbf{3P\%} & \textbf{DNT} & \textbf{Block 3P} & \textbf{uBlock} & \textbf{GPC} \\
\midrule

\multicolumn{8}{@{}l}{\cellcolor{gray!15}\textbf{\quad CCPA Subject (552 sites, 11,553 cookies)}} \\
\addlinespace[1pt]

\cellcolor{gray!20}Targeting          & 3,670 & 51.0 & 4.1 & \gradientcell{34}{0}{100}{red}{green}{50} (8)  &
\gradientcell{28}{0}{100}{red}{green}{50} (1)  & \gradientcell{83}{0}{100}{red}{green}{50} (87) & \gradientcell{55}{0}{100}{red}{green}{50} (41) \\
Performance        & 2,038 & 66.5 & 6.1 & \gradientcell{14}{0}{100}{red}{green}{50} (9)  & \gradientcell{6}{0}{100}{red}{green}{50} (3)   &
\gradientcell{79}{0}{100}{red}{green}{50} (83) & \gradientcell{38}{0}{100}{red}{green}{50} (35) \\
\cellcolor{gray!20}Functional         & 1,431 & 41.6 & 5.6 & \gradientcell{10}{0}{100}{red}{green}{50} (9)  &
\gradientcell{5}{0}{100}{red}{green}{50} (1)   & \gradientcell{47}{0}{100}{red}{green}{50} (71) & \gradientcell{30}{0}{100}{red}{green}{50} (30) \\
Strictly Necessary &   1,018 & 44.2 & 5.0 & \gradientcell{18}{0}{100}{red}{green}{50} (5)  & \gradientcell{14}{0}{100}{red}{green}{50} (7)  &
\gradientcell{46}{0}{100}{red}{green}{50} (48) & \gradientcell{27}{0}{100}{red}{green}{50} (18) \\
\cellcolor{gray!20}Unknown            & 3,396 & 10.3 & 2.7 & \gradientcell{57}{0}{100}{red}{green}{50} (27) &
\gradientcell{56}{0}{100}{red}{green}{50} (26) & \gradientcell{68}{0}{100}{red}{green}{50} (73) & \gradientcell{64}{0}{100}{red}{green}{50} (47) \\

\cmidrule{1-8}

\textit{Subtotal} & \textit{11,553} & \textit{40.0} & \textit{4.3} & \textit{33 (15)} & \textit{28 (9)} & \textit{70 (82)} & \textit{49 (37)} \\

\addlinespace[3pt]
\multicolumn{8}{@{}l}{\cellcolor{gray!15}\textbf{\quad CCPA Not-Subject (363 sites, 7,566 cookies)}} \\
\addlinespace[1pt]

\cellcolor{gray!20}Targeting          & 2,722 & 45.0 & 5.0 & \gradientcell{44}{0}{100}{red}{green}{50} (23) &
\gradientcell{41}{0}{100}{red}{green}{50} (20) & \gradientcell{93}{0}{100}{red}{green}{50} (93) & \gradientcell{56}{0}{100}{red}{green}{50} (34) \\
Performance        & 1,426 & 78.3 & 5.8 & \gradientcell{10}{0}{100}{red}{green}{50} (2)  & \gradientcell{6}{0}{100}{red}{green}{50} (1)   &
\gradientcell{87}{0}{100}{red}{green}{50} (88) & \gradientcell{27}{0}{100}{red}{green}{50} (18) \\
\cellcolor{gray!20}Functional         &   663 & 53.5 & 6.9 & \gradientcell{12}{0}{100}{red}{green}{50} (8)  &
\gradientcell{6}{0}{100}{red}{green}{50} (3)   & \gradientcell{57}{0}{100}{red}{green}{50} (66) & \gradientcell{22}{0}{100}{red}{green}{50} (14) \\
Strictly Necessary &   534 & 49.8 & 5.1 & \gradientcell{23}{0}{100}{red}{green}{50} (16) & \gradientcell{21}{0}{100}{red}{green}{50} (14) &
\gradientcell{68}{0}{100}{red}{green}{50} (71) & \gradientcell{24}{0}{100}{red}{green}{50} (20) \\
\cellcolor{gray!20}Unknown            & 2,221 &  6.6 & 1.6 & \gradientcell{79}{0}{100}{red}{green}{50} (64) &
\gradientcell{79}{0}{100}{red}{green}{50} (63) & \gradientcell{86}{0}{100}{red}{green}{50} (86) & \gradientcell{81}{0}{100}{red}{green}{50} (73) \\

\cmidrule{1-8}

\textit{Subtotal} & \textit{7,566} & \textit{41.1} & \textit{4.3} & \textit{43 (18)} & \textit{41 (14)} & \textit{85 (89)} & \textit{53 (31)} \\

\midrule
\textbf{Total (915 sites)} & \textbf{19,119} & \textbf{40.4} & \textbf{4.3} & \textbf{37 (17)} & \textbf{33 (11)} & \textbf{76 (85)} & \textbf{51
(34)} \\

\bottomrule
\end{tabular}
} 

\vspace{1mm}
\raggedright
\scriptsize
\end{table}

%% file: tables/top-cookie-scripts.tex
    \begin{table}[!t]
    \centering
    \caption{Top 10 third-party scripts responsible for setting 58.4\% of Targeting cookies in the first-party context.}
    \label{tab:top-cookie-scripts}
    \footnotesize
    \setlength{\tabcolsep}{3pt}
    \begin{tabular}{p{3.3cm}rrrc}
    \toprule
    \textbf{Script Domain} & \textbf{Cookies} & \textbf{Sites} & \textbf{\% Sample} & \textbf{Category} \\
    \midrule
    \cellcolor{gray!20}securepubads.g.doubleclick.net
        & 294 & 144 & \gradientcell{15.7}{0}{60}{red}{green}{40}\% & Advertising \\
    www.googletagmanager.com
        & 278 & 239 & \gradientcell{26.1}{0}{60}{red}{green}{40}\% & Tag Mgmt. \\
    \cellcolor{gray!20}bat.bing.com
        & 273 & 132 & \gradientcell{14.4}{0}{60}{red}{green}{40}\% & Ads/Analytics \\
    connect.facebook.net
        & 196 & 192 & \gradientcell{21.0}{0}{60}{red}{green}{40}\% & Social/Ads \\
    \cellcolor{gray!20}assets.adobedtm.com
        & 181 & 59 & \gradientcell{6.4}{0}{60}{red}{green}{40}\% & Tag Mgmt. \\
    analytics.tiktok.com
        & 131 & 62 & \gradientcell{6.8}{0}{60}{red}{green}{40}\% & Social/Analytics \\
    \cellcolor{gray!20}ak.sail-horizon.com
        & 120 & 43 & \gradientcell{4.7}{0}{60}{red}{green}{40}\% & Marketing Auto. \\
    tags.tiqcdn.com
        & 119 & 34 & \gradientcell{3.7}{0}{60}{red}{green}{40}\% & Tag Mgmt. \\
    \cellcolor{gray!20}sc-static.net
        & 117 & 44 & \gradientcell{4.8}{0}{60}{red}{green}{40}\% & Audience Meas. \\
    cdn.attn.tv
        & 100 & 18 & \gradientcell{2.0}{0}{60}{red}{green}{40}\% & Marketing \\
    \bottomrule
    \end{tabular}
  \end{table}

%% file: 5-discussion.tex
\section{Discussion}
\label{sec:discussion}
\myparagraph{The impact of the CCPA on privacy practices} 
Our results show that many \notsubject websites provide high quality privacy disclosures, explicitly referencing the CCPA despite lacking clear legal obligations to do so. This spillover suggests that the statute exerts influence over privacy communication beyond its formal scope. We hypothesize that this ripple effect is partially driven by the difficulty of precisely determining CCPA applicability, as observed in both our work and prior work in gathering website subjectivity information~\cite{tran2024measuring}. Given the ambiguities in scope, firms may strategically over-comply to mitigate risk. While this arguably raises the baseline for user transparency, it also imposes non-trivial compliance costs on organizations and complicates enforcement. This problem is exacerbated by a fragmented landscape where multiple state-level privacy laws with unique applicability criteria co-exist. 

We also find that both \subject and \notsubject websites engage in cookie writes, especially from third-parties, at comparable levels. This finding is partly consistent with the CCPA's design: the statute regulates the \textit{sale or sharing} of user information and opt-out honoring, not cookie-writes themselves. However, cookie writes are a necessary \textit{precondition} for privacy practices regulated by the CCPA. Our findings highlighting the widespread deployment of third-party targeting cookies and the incomplete response to opt-out signals raise a critical question: \textit{Should the CCPA expand its legal definition to granular data collection practices, including cookies, mirroring the GDPR?} We argue in the affirmative, since such regulation could (1)~offer meaningful benefits to users by preventing the initial layer of data collection, (2)~encourage greater front-end transparency from websites, and (3)~enable faster third-party auditing of non-compliance, such as through \name. 

\myparagraph{Multi-Layer Auditing} Combining auditing of disclosures with tracking behavior allows \name to surface audit-relevant patterns, such as websites that claim to honor GPC while continuing to set third-party cookies when the GPC is set. Such signals are useful precisely because regulators and researchers cannot manually inspect every website, policy, consent interface, and tracking configuration at scale. \name's outputs do not replace legal analysis and enforcement, but can inform audit selection, public reporting, rulemaking, and future enforcement priorities. We make \name open-source (see Appendix~\ref{sec:open-science}) and are actively engaging with regulators and legal experts to explore use-cases in regulatory auditing. 
In addition, we also believe \name can be useful as a self-auditing tool for website developers and privacy teams as it can translate dense legal requirements into structured policy checks and browser measurements.

\myparagraph{Implications Beyond the CCPA}
Although our analysis focuses on the CCPA, the gaps we document are unlikely to be unique to California. More than 20 U.S.\ states have enacted CCPA-like privacy laws~\cite{iapp-us-privacy-laws-2025}, many with similar opt-out frameworks and universal opt-out signals such as GPC. The 2025 multi-state GPC enforcement sweep by California, Colorado, and Connecticut~\cite{cppa_gpc_sweep_2025} further suggests that regulators view incomplete opt-out honoring as a shared concern. 
Future research can extend \name's rubric and crawling infrastructure to newer state laws~\cite{va_cdpa_rights,co_cpa_scope, ut_ucpa_rights}.


\myparagraph{Limitations} 
While \name automates much of our policy and cookie analysis, it still requires manual effort to engineer and validate LLM prompts, which remains necessary to reduce hallucinations. Although prior work shows that internal pages can differ from landing pages in resource composition and tracking behavior~\cite{aqeel2020landing}, we verify that homepages capture roughly 87--90\% of the third-party Targeting trackers on a typical internal page across all six configurations, so homepage-only measurement does not severely undercount tracking (Appendix~\ref{app:innerpage}). Our measurements nonetheless remain limited to cookies observable without authenticated or interaction-gated flows (e.g., login, scrolling), which we leave to future work. Additionally, our analysis focuses on cookie-based tracking, which is prominent but only one mechanism for user data collection. \name can be extended to other user-facing techniques such as browser fingerprinting~\cite{eckersley2010unique,acar2014web} and CNAME cloaking~\cite{dimova2021cname} that can complement or bypass cookie-based tracking in future work.
Our results represent a single crawl conducted in mid-2025, and cookie-setting behavior may vary across sessions due to temporal changes in tracking infrastructure. However, prior longitudinal studies have found that both the structure of the tracking ecosystem and GPC compliance rates remain relatively stable over short periods~\cite{solomos2020clash,hausladenwebsites}. Despite these limitations, we believe \name provides a strong foundation for repeatable and automated auditing of privacy practices.

%% file: 99-appendix.tex
\appendix
\section{Generative AI Usage}
Generative AI tools (\eg Claude) were used for grammar checking and improving readability of the manuscript. Additionally, as described in Section~\ref{sec:llm-policy-scoring}, our methodology uses LLMs (GPT-3.5, GPT-4o, and Gemini 2.5) for automated privacy policy evaluation. All content was written, reviewed, and verified by the authors.

\section{Open Science}
\label{sec:open-science}
All artifacts necessary to evaluate the contributions of this paper are available at \url{https://github.com/r-andlab/PrivAudit}. The repository contains:
\begin{itemize}
\item \textbf{Cookie crawler:} Puppeteer-based crawler with six privacy configuration profiles and per-website cookie banner flow definitions (\texttt{crawler/}).
\item \textbf{Privacy policy scraper:} Selenium-based pipeline for automated discovery and extraction of privacy policy text from website homepages (\texttt{policy\_analysis/scraper/}).
\item \textbf{LLM policy analysis:} Analysis notebook, statistical significance tests, and the CCPA-specific rubric and structured LLM prompt used for policy evaluation (\texttt{policy\_analysis/}; Appendix~\ref{sec:full-llm-prompt}).
\item \textbf{Cookie categorization:} Multi-source classification pipeline (\texttt{cookie\_categorization/}).
\item \textbf{Banner detection:} Consent banner detection module (\texttt{banner\allowbreak\_detection/}).
\item \textbf{Analysis scripts:} All table and figure generation scripts for reproducing the paper's results (\texttt{analysis/}).
\item \textbf{Datasets} (\texttt{data/data.zip}): Master cookie dataset, LLM policy audit results, website classification labels, cookie data for websites with consent banners, and scraped privacy policy text files.
\end{itemize}

\noindent\textbf{Artifacts not included:} The cookie categorization pipeline requires raw third-party cookie classification databases (Cookiepedia, Open Cookie Database, Cookie Cutter DB) that are not included directly due to licensing. Download instructions from original sources are provided in the repository README. The categorization results are already embedded in the master dataset.

\section{Ethical Considerations}
\label{sec:ethics}
We conducted all crawling and analysis from our clients located in California. \name makes only a handful of requests to each public website: (1) to retrieve its privacy policy and (2) to observe the cookies set under multiple privacy configurations. In total, we issued fewer than 10 requests per site, an insignificant fraction of normal web traffic. To select internal pages, we relied on each site's own published sitemap (the \texttt{Sitemap} directive in its \texttt{robots.txt}), or a small set of common public paths when none was available. The crawls under different privacy configurations were performed in a round-robin manner to avoid any load on the websites. We did not collect or interact with any user data other than our own, and all analysis, including data fed to LLMs, relied solely on publicly available website responses.


\section{Website Categorization}
\label{appendix:domain-categorization}
 \input{tables/accessible-websites-categories}
We provide a breakdown on the various website categories derived from Tran~\etal~\cite{tran2024measuring} in Table~\ref{tab:accessible-websites-categories}. We consider only for-profits clearly meeting the CCPA applicability criteria as \subject. All other categories are \notsubject according to CCPA guidelines.

\section{Scaling to Unlabeled Websites}
\label{app:tranco}

The Popular-Website list---the 1,000 unlabeled websites in our scaling case study---was selected from the Tranco Top-1M permanent list~\cite{Tranco} by the following protocol: (1) keep one base domain per site under generic or U.S.-oriented TLDs (\texttt{.com}, \texttt{.org}, \texttt{.net}, \texttt{.us}, \texttt{.io}, \texttt{.co}); (2) require an active homepage (HTTP~200, \texttt{text/html}), which removes non-consumer infrastructure such as CDN, DNS, and API endpoints; and (3) draw 250 sites per rank band (top-1k, 1k--10k, 10k--100k, 100k--1M) at random, excluding every domain in the Known-Subjectivity list. We then run the same GPT-4o policy-scoring pipeline as our main analysis and the cookie audit under the six browser-native privacy configurations. Policies we could not retrieve fall outside the scope of our analysis.

\myparagraph{Results} Disclosure on this corpus is weaker than on the Known-Subjectivity list: only 72\% of sites expose a usable privacy policy (versus 93\%), and those that do score lower on completeness, usability, and accuracy (1.65, 1.78, and 1.94 out of 3, versus 2.33, 2.29, and 2.43 for \subject websites) and less often disclose the rights to access (40\%), delete (36\%), and opt out (52\%) than \subject websites (85\%, 86\%, and 77\%); only 28\% present a machine-detectable CCPA opt-out (``Do Not Sell'' / ``Your Privacy Choices'') mechanism. Tracking, by contrast, is more pervasive: the 740 sites that set any cookie place 7{,}244 in total at initial load (median 6 per site), of which \textit{1{,}650 (22.8\%) are Targeting}, and 67\% of all cookies (79\% of Targeting cookies) are set by third-party scripts, more than the 40\% on the Known-Subjectivity list (Table~\ref{tab:comprehensive-third-party-analysis}). Privacy protections follow the same pattern as on the Known-Subjectivity list: uBlock Origin removes 96\% of third-party Targeting cookies and GPC reduces them by 22\% ($p<0.001$), while DNT, third-party-cookie blocking, and Consent-O-Matic show no measurable effect. The disclosure--behavior gap also appears here: among the sites whose policies claim not to sell personal data, \textit{43.6\% (99 sites) still set at least one third-party Targeting cookie at initial load}, aligning with the 50.8\% we report for \subject websites in \S\ref{sec:policy-behavior-results}, and of the sites that present an opt-out mechanism, 54\% still set a third-party Targeting cookie. These results confirm that \name's audit signals can extend to arbitrary corpora of websites.

\takeaway{On the Popular-Website list (1,000 sites), privacy disclosures are weaker yet the disclosure--behavior gap persists: 43.6\% of sites that claim not to sell data still set third-party Targeting cookies at first visit.}

\section{Cookie Writing by Consent Flow}
\input{figures/new_figures/consent_flow_cookies_ccpa}

Consent-banner rejection provides only limited protection because many cookies are set before or despite consent interaction.  Among the 239 websites with detected cookie banners, accepting cookies increases the median cookie count from 23 to 27 for \subject websites and from 15 to 16 for \notsubject websites (distribution in Figure~\ref{fig:consent-flow-cookies-ccpa}). Rejecting cookies, however, produces only a modest reduction for \subject websites (23 to 22 cookies) and no change for \notsubject websites (15 to 15 cookies), reflecting sites that set few cookies initially but add cookies after any consent interaction. Across all consent stages, \subject websites set roughly 50\% more cookies than \notsubject websites. 

\takeaway{User consent actions have limited effect on cookie collection behavior.}

\section{Cookie Security and Privacy Characteristics} 
\label{para:cookie-characteristics}
\myparagraph{Security Attributes} Table~\ref{tab:cookie-attributes-party} shows the percentage of cookies that adopt various security attributes such as \texttt{Secure} and \texttt{SameSite}. \textit{We observe that only 25.4\% of all cookies use the \texttt{Secure} attribute, with significant differences between \subject (29.1\%) and \notsubject (19.8\%) websites ($\chi^2$, $p<0.001$, $V=0.10$), and this percentage is particularly low for Targeting (20.2\%) and Performance cookies (23.1\%)}, which may contain personally identifiable information.  Furthermore, the vast majority of cookies can be accessed through Javascript, as evidenced by the low adoption (5.1\%) of the \texttt{HttpOnly} attribute across all websites, though \subject websites show slightly higher adoption (6.4\%) than \notsubject websites (3.0\%, $p<0.001$, $V=0.08$). The adoption of the \texttt{SameSite} attribute is also relatively low (16.1\%) across both \subject and \notsubject websites, with no practically meaningful difference between groups ($V=0.02$). We observe that a majority of cookies (57.1\%) are set in a persistent manner rather than as session cookies, with only 39.5\% of Targeting cookies set as session cookies. We see that Targeting cookies are especially long-lived, as shown in Figure~\ref{fig:app-cookie-longevity}, with a median lifespan of almost a year, whereas other types of cookies have a smaller lifespan. These findings suggest that websites prioritize long-term user tracking rather than session-based tracking, an issue exacerbated by the limited disclosure of data retention practices in privacy policies (\S\ref{sec:privacy-policy-results}).
\input{tables/cookie-attributes-party}

\myparagraph{Geographic and Network Information in Cookies} We examined cookie values for potential personal information (as defined by the GDPR) stored in plaintext, focusing on validated instances of geographic and network data. \textit{We identified 363 cookies (1.9\% of total) with location-related names such as \texttt{geoData}, \texttt{zipcode}, and \texttt{geo}, spanning 171 unique cookie identifiers.} Within these cookies, we detected 7 instances of latitude/longitude coordinates, 6 ZIP codes, and 226 IP addresses. Notably \textit{162 IP addresses appear in cookies classified as ``Functional'' and 28 in ``Strictly Necessary'' cookies}, raising questions about the interpretation of necessity under CCPA. We note that this is a lower bound of personal information as we only analyze plaintext information in cookies. 

Privacy-enhancing configurations demonstrate limited effectiveness in removing this sensitive information. \textit{GPC fails to reduce geographic PII, with ZIP code instances remaining constant (6 instances) and latitude/longitude coordinates actually increasing from 7 to 8 instances.} IP addresses show modest reduction under GPC (15\%, from 226 to 193 instances) and uBlock Origin (14\%, from 226 to 195 instances). This persistence of location and network identifiers despite opt-out signals suggests that current privacy mechanisms inadequately address geolocation-based tracking.

\section{Delegation Through Nested Scripts} 
We observe that third-party cookie control frequently manifests through nested script inclusion chains. Tag management systems such as Google Tag Manager (GTM) serve as intermediaries, loading dozens of additional third-party tracking scripts that subsequently write cookies nominally scoped to the publisher's domain. Using our call-stack instrumentation, we observed an average of 8.2 function frames in cookie-setting traces, typically reaching depths of 40-50 frames, with extreme cases exceeding 500 layers of delegation across tag managers, ad networks, and analytics frameworks. These delegation chains enable third-party scripts to set cookies with first-party domain scopes, thereby evading browser-imposed restrictions on third-party cookie access while maintaining cross-site tracking capabilities. A representative example observed on major news websites shows the \texttt{\_ga} Google Analytics cookie—carrying a first-party domain scope—being set through a delegation chain originating from the site's own tag manager, which loads \texttt{googletagmanager.com/gtm.js}, which in turn loads \texttt{google-analytics.com/analytics.js} that finally executes the cookie write.

\myparagraph{Supply-Chain Concentration and Enforcement Implications} 
Our analysis reveals that third-party cookie deployment is dominated by a small number of platforms: just 10 scripts from Google, Microsoft, Meta, and Adobe account for 58\% of all script-set targeting cookies. This concentration suggests that improving privacy behavior may require attention not only to individual websites as in prior cases~\cite{sephora_ccpa_2022,tractor_supply_cppa}, but also to the upstream services that deploy and manage tracking infrastructure. If major tag managers, advertising scripts, or analytics SDKs honored opt-out signals more consistently by default, the effect could propagate across many websites at once. Our script attribution exposes a layer of the tracking supply chain that is often hidden from domain-based cookie analysis performed by browsers, and can therefore support both technical defenses and regulatory oversight.

Our findings also reinforce the importance of recent work on first-party cookie defenses. CookieGuard~\cite{bahrami2025cookieguard} and CookieGraph~\cite{munir2023cookiegraph} show that most first-party tracking cookies are written by third-party scripts, and propose mechanisms to isolate this behavior. We extend this line of work by showing that these script-set cookies are also more resistant to privacy signals like GPC.

\section{Cookie Longevity}
\label{sec:cookie-lifetimes}

\input{figures/app-cookie-longevity}
Figure~\ref{fig:app-cookie-longevity} shows the lifespan of various types of cookies set across \subject and \notsubject websites. We observe that Targeting and Performance cookies generally tend to be more long-lived, with $\sim$50\% of targeting cookies having lifetimes of almost a year. Overall, around 23.5\% of cookies are short-lived (0-1 days) which are dominated by Strictly Necessary and Functional cookies, and 32.7\% of cookies have a lifetime between 1-2 years. Both \subject and \notsubject websites exhibit similar cookie longevity, with a median lifetime of roughly one year for each group.

\section{Policy and Cookie Analysis by Industry Sector}
\label{sec:app-industry}
\input{figures/appendix-industry-policy}
\myparagraph{Industry Trends in Privacy Policy Analysis} Figure~\ref{fig:appendix-industry-policy} shows the percentage of policies by industry sector that discuss behavioral claims and disclosures.  The spillover of CCPA compliance to \notsubject websites is driven primarily by for-profit industries, which demonstrate substantially higher disclosure coverage and behavioral claims compared to Government and non-profit entities. While over 80\% of for-profit websites disclose users' rights to access and delete personal data, Government websites show dramatically lower rates (7\% and 4\%, respectively). This disparity is even more pronounced for privacy signals: approximately 23\% of for-profit websites claim to honor GPC, while no Government websites do so, highlighting that voluntary CCPA adoption remains concentrated in the commercial sector.

\myparagraph{Industry Trends in Cookie-based Tracking} We find that website categories vary significantly in cookie deployment strategies. Targeting cookies are especially prevalent in for-profit websites (33.8\% of all cookies) compared to non-profit and government sites (24.8\%), regardless of CCPA applicability. This disparity is particularly notable given that non-profit and government sectors provide the least detailed privacy policies, as discussed earlier. Within for-profit industries, those most reliant on advertising and audience tracking---such as Automotive (17.5 targeting cookies per site on average), Beauty (14.0 per site), and News Media (10.8 per site)---demonstrate substantially elevated targeting footprints.

\section{Internal-Page vs.\ Homepage Tracking}
\label{app:innerpage}
\input{tables/inner-page}
To validate that homepage measurement is representative (\S\ref{sec:tracking-results}), we crawled up to five internal pages per site under all six privacy configurations (Default, GPC, DNT, third-party-cookie blocking, uBlock Origin, and Consent-O-Matic). Internal pages are sampled from the same-site links on the rendered homepage (excluding the homepage, static assets, and logout links, de-duplicated by path), drawing up to five at random with a per-site seed for reproducibility, following prior internal-page measurement studies~\cite{aqeel2020landing}. Each page is loaded in a fresh, isolated browser context with the same script-based third-party attribution and categorization as our main crawl.

Table~\ref{app:innerpage-tab} reports, per configuration, the per-site paired difference between internal-page and homepage third-party Targeting cookies and the fraction of a site's distinct third-party Targeting trackers already present on the homepage. Across all six configurations, internal pages set significantly fewer third-party Targeting cookies than homepages. We report two measures: the fraction of a \emph{typical internal page's} distinct third-party Targeting trackers already present on the homepage (87--90\%), and the stricter fraction of the \emph{union} of trackers across all internal pages captured by the homepage (78--83\%). The gap between the two reflects that internal pages collectively surface a tail of additional trackers, but such tail third parties are largely non-deterministic between repeated visits~\cite{urban2020beyond}, so the per-page measure better reflects representativeness. Our findings are consistent with prior studies that  show that individual internal pages carry fewer trackers than landing pages~\cite{aqeel2020landing}. 
Under uBlock Origin, targeting cookies are nearly eliminated on both homepages and internal pages, so its paired difference is near zero by construction.


\section{Firm-Size Controls and Multiple-Testing Correction}
\label{app:size-control}
\input{tables/firm-size}
CCPA-subject firms are, by the statute's revenue threshold, larger on average and may derive more revenue from data sharing, which could on its own lead to more detailed privacy policies. To test whether the disclosure gap reflects CCPA \emph{applicability} rather than size or ad-revenue reliance, we fit a logistic regression for each key disclosure on CCPA subjectivity, adjusting for company annual revenue ($\log_{10}$, from the business-intelligence source used for subjectivity in \S\ref{sec:website-selection}) and a proxy for ad-revenue reliance (each site's third-party Targeting tracking intensity), over the $N=816$ sites with available revenue data (Table~\ref{tab:size-control}, panel A). CCPA subjectivity remains a strong, positive predictor of the opt-out, access, and deletion disclosures (all $p<0.001$), whereas company revenue is not a significant predictor of the opt-out ($p=0.55$) and is at most weakly associated with the other disclosures.

We complement this with a size-matched comparison restricted to organizations above the statute's \$25M revenue threshold ($N=563$: 528 subject, 35 exempt), contrasting \subject firms with large but exempt organizations such as nonprofits and government agencies (Table~\ref{tab:size-control}, panel B). Even at comparable size, \subject firms disclose opt-out mechanisms far more often (77\% vs.\ 46\%), with comparable gaps for the rights to access and delete data (85\% vs.\ 60\%). Together, these results indicate that the disclosure gap reflects CCPA applicability rather than firm size or ad-revenue reliance alone, though as an observational comparison it cannot establish strict causation. Table~\ref{app:mht} further reports multiple-testing--adjusted $p$-values for all 25 tests; every main policy finding survives the stricter Holm--Bonferroni correction.

\section{Multiple Hypothesis Corrections}
\label{sec:multiple-hypothesis}
\input{tables/corrections}
To account for multiple comparisons which we perform in \S\ref{sec:results}, we treat all 25 \subject versus \notsubject tests (16 policy and 9 cookie) as a single family and report both Benjamini--Hochberg and  Holm--Bonferroni adjusted $p$-values in Table~\ref{app:mht}. We base our conclusions on the stricter Holm correction, under which all main policy findings remain significant.

\section{Legal Basis and Mapping}
\label{sec:legal-basis}

\input{tables/legal-basis}
The California Consumer Privacy Act (CCPA) and its amendment, the California Privacy Rights Act (CPRA), establish explicit disclosure obligations and consumer rights that businesses must incorporate into their privacy policies. To ensure that our structured LLM prompt aligns with these legal requirements, we map each field in the schema to the corresponding statutory provision, which we then validate through an expert legal review. Table~\ref{tab:ccpa_legal_mapping} presents the detailed alignment between our prompt fields and the relevant CCPA/CPRA sections.

\myparagraph{Legal Validation of the Rubric}
The alignment in Table~\ref{tab:ccpa_legal_mapping} was validated through a structured expert-review process. The authors first manually constructed an initial mapping of each CCPA/CPRA provision to a corresponding compliance dimension in our rubric. Then, the mapping was reviewed by an external legal expert specializing in California privacy regulation, an attorney holding a Juris Doctor (JD) with over 15 years of experience in privacy law and regulatory compliance spanning both
industry privacy programs and government privacy oversight, currently serving at a public-sector regulatory organization. The expert reviewed each provision-to-dimension mapping explicitly, confirmed that the alignment correctly reflected statutory requirements, and suggested minor interpretive refinements, which we incorporated: (i)~for sensitive personal information, that regulated entities must additionally ensure reasonable security of such data (\S1798.100(e)); and (ii)~for the deletion right, that it is
subject to limited statutory exceptions (e.g., completing a transaction or complying with a legal obligation, \S1798.105(d)) and that businesses must respond within 45 days (\S1798.130(a)(2)(A)).
In addition, one of the co-authors has worked directly with government agencies on privacy regulations and contributed to continuous legal interpretation. This process validates the legal rubric itself, that each dimension correctly reflects the statute, and not the LLM application of the rubric, which is impractical to be expert-validated at scale. The correctness of the LLM's
\emph{application} of the rubric is instead evaluated separately through manual output review and inter-model agreement (\S\ref{sec:llm-validation}).

\section{Full LLM Prompt}
\label{sec:full-llm-prompt}

For transparency and reproducibility, we provide the complete instruction prompt that guided the large language model (LLM) in our study below. This prompt encodes both a normative framework derived from the CCPA, and a practical data-extraction schema that allows for large-scale, automated, and reproducible auditing of privacy policies.
The prompt is designed with three guiding principles: (1) \emph{statutory alignment}, meaning every field corresponds directly to a disclosure or right enumerated in the CCPA; (2) \emph{structured reproducibility}, meaning outputs conform to a single fixed JSON schema with deterministic key ordering, strict types, and no extraneous content; and (3) \emph{evidence-based interpretability}, meaning every non-trivial flag or score must be accompanied by textual justification grounded in the policy itself. These principles ensure that outputs are both legally meaningful and auditable by human reviewers.

The first two fields, \texttt{online\allowbreak\_data\allowbreak\_practices} and \texttt{offline\allowbreak\_data\allowbreak\_practices}, capture concise natural-language summaries of how the policy describes data collection channels. This dual distinction reflects the statutory scope of the CCPA, which covers information collected “online and offline” (\S1798.140(v)(1)). Online practices include cookies, SDKs, analytics tools, ad-tech integrations, or registration forms embedded in websites or apps. Offline practices capture in-store purchases, customer service interactions, loyalty programs, events, or any physical-world collection. By requiring explicit summaries, and defaulting to ``Not mentioned.'' when absent, the schema prevents silent omissions.

The central block, \texttt{rubric\_assessment}, quantitatively evaluates the policy’s presentation of consumer rights and disclosure obligations. It includes:  
\begin{itemize}
  \item \emph{completeness\_score (0–3)}: Does the policy enumerate core statutory rights---access (\S1798.110), deletion (\S1798.105), opt-out of sale or sharing (\S1798.120), and (under CPRA) correction (\S1798.106) and sensitive PII limitation (\S1798.121)? Does it explain how to exercise them? A score of 0 reflects total absence; 3 indicates comprehensive coverage with actionable mechanisms.  
  \item \emph{usability\_score (0–3)}: How easy are the mechanisms to use? We assess the availability of multiple clear channels (e.g., toll-free number, web portal, email), the presence of explicit labels (e.g., ``Do Not Sell or Share My Personal Information''), guidance on verification, timelines for responses, and any references to appeal processes. This corresponds to procedural requirements in \S1798.130(a)(1).  
  \item \emph{accuracy\_score (0–3)}: Are statutory terms used correctly and up to date? For example, policies must distinguish ``selling'' from ``sharing'' for cross-context behavioral advertising, acknowledge opt-out signals such as Global Privacy Control, and avoid outdated claims such as ``we do not honor CCPA rights.'' Misstatements reduce the score.  
\end{itemize}
Each score is paired with a mandatory free-text rationale, ensuring that numeric judgments are traceable to policy text. This block also includes a \texttt{policy\_contradiction} flag to capture internal inconsistencies (e.g., a policy that both claims to ``never share'' and later describes sharing), and a \texttt{disclosure\_map} of booleans marking whether the policy explicitly discloses each statutory element: categories collected (\S1798.130(a)(5)\allowbreak(B)(i)), categories of sources (\S1798.130(a)(5)\allowbreak(B)(ii)), purposes of collection (\S1798.130(a)(5)\allowbreak(B)(iii)), categories of third parties (\S1798.130(a)(5)\allowbreak(B)(iv)), retention periods (\S1798.100(a)(3)), and rights (access, delete, opt-out). By binarizing disclosures, we ensure comparability across policies.

\myparagraph{Behavioral Claims}  
While disclosures capture what a policy states, they may not reveal how the business actually behaves. To bridge this gap, the schema includes a \texttt{behavioral\_\allowbreak-claims} block. Each claim is expressed as \texttt{true}, \texttt{false}, or \texttt{``unspecified''}, covering:  
\begin{itemize}
  \item signal recognition (honors GPC, respects DNT),  
  \item cookie and consent handling (sets cookies before consent, after rejection, deletes cookies upon rejection, uses tracking only after consent), and  
  \item data transfer practices (sells data, shares with third parties, shares for cross-context behavioral advertising).  
\end{itemize}
Every flag must be accompanied by a justification quoting or paraphrasing the nearest relevant clause, or—if absent—an explicit statement that ``No explicit statement found; flag set to unspecified.'' This ensures that unspecified values are not silent defaults but traceable to an absence of evidence. This design makes the schema robust to both over-claiming (hallucinated compliance) and under-claiming (false negatives).

\myparagraph{Cookies and Statutory Mentions}  
Because cookies are a central mechanism of online tracking, the schema includes dedicated fields for whether cookies are mentioned (\texttt{cookie\allowbreak\_practices\allowbreak\_mentioned}) and for an excerpt summarizing the discussion (\texttt{cookie\allowbreak\_practices\allowbreak\_excerpt}). This ensures that cookie-related practices are consistently identified even if not linked to explicit rights. A parallel field captures whether the policy explicitly names the CCPA or CPRA, and if so, records a concise excerpt (\texttt{mentions\allowbreak\_ccpa}, \texttt{mentions\allowbreak\_ccpa\allowbreak\_excerpt}). These allow us to distinguish between policies that acknowledge California-specific legal frameworks and those that do not.

Table~\ref{tab:ccpa_legal_mapping} provides the full mapping of each of our LLM-prompt fields with specific CCPA guidelines. By enforcing a schema that is tightly mapped to statute, the LLM output is not only machine-parseable but also legally interpretable.

\clearpage
\begingroup
\onecolumn
\begin{lstlisting}[language=Python,breaklines=true,basicstyle=\ttfamily\footnotesize,columns=fullflexible]

def build_prompt(domain: str, policy_text: str) -> str:
    """
    Returns a detailed instruction prompt for the LLM with clear scoring rubrics,
    definition of each metric/flag, and a strict JSON schema to follow.
    Also instructs to always provide non-empty justifications.
    """
    return f"""
You are a privacy compliance analyst evaluating a privacy policy strictly under the California Consumer Privacy Act (CCPA) and related U.S. state privacy concepts (CPRA updates included). You must rely ONLY on the policy text provided below--no external knowledge or assumptions.

Analyze the policy for: {domain}

# Output Requirements
Return one JSON object matching this exact schema and types:
{
  "online_data_practices": "string (brief summary or 'Not mentioned.')",
  "offline_data_practices": "string (brief summary or 'Not mentioned.')",
  "rubric_assessment": {
    "completeness_score": 0-3,
    "completeness_reason": "string",
    "usability_score": 0-3,
    "usability_reason": "string",
    "accuracy_score": 0-3,
    "accuracy_reason": "string",
    "disclosure_map": {
      "data_collected": true/false,
      "data_shared": true/false,
      "purpose_of_collection": true/false,
      "retention_period": true/false,
      "right_to_access": true/false,
      "right_to_delete": true/false,
      "opt_out": true/false
    }
  },
  "behavioral_claims": {
    "honors_gpc": true/false/"unspecified",
    "respects_dnt": true/false/"unspecified",
    "sets_cookies_before_consent": true/false/"unspecified",
    "sets_cookies_after_rejecting_consent": true/false/"unspecified",
    "deletes_cookies_on_rejection": true/false/"unspecified",
    "uses_tracking_only_after_consent": true/false/"unspecified",
    "sells_data": true/false/"unspecified",
    "shares_with_third_parties": true/false/"unspecified",
    "justifications": {
      "honors_gpc": "1-2 sentences. NEVER 'unspecified'. If flag is 'unspecified', explicitly say 'No explicit statement found; flag set to unspecified.' and cite the nearest relevant line(s).",
      "respects_dnt": "1-2 sentences. NEVER 'unspecified'. If flag is 'unspecified', explicitly say 'No explicit statement found; flag set to unspecified.' and cite the nearest relevant line(s).",
      "sets_cookies_before_consent": "1-2 sentences. NEVER 'unspecified'. If flag is 'unspecified', explicitly say 'No explicit statement found; flag set to unspecified.' and cite the nearest relevant line(s).",
      "sets_cookies_after_rejecting_consent": "1-2 sentences. NEVER 'unspecified'. If flag is 'unspecified', explicitly say 'No explicit statement found; flag set to unspecified.' and cite the nearest relevant line(s).",
      "deletes_cookies_on_rejection": "1-2 sentences. NEVER 'unspecified'. If flag is 'unspecified', explicitly say 'No explicit statement found; flag set to unspecified.' and cite the nearest relevant line(s).",
      "uses_tracking_only_after_consent": "1-2 sentences. NEVER 'unspecified'. If flag is 'unspecified', explicitly say 'No explicit statement found; flag set to unspecified.' and cite the nearest relevant line(s).",
      "sells_data": "1-2 sentences. NEVER 'unspecified'. If flag is 'unspecified', explicitly say 'No explicit statement found; flag set to unspecified.' and cite the nearest relevant line(s).",
      "shares_with_third_parties": "1-2 sentences. NEVER 'unspecified'. If flag is 'unspecified', explicitly say 'No explicit statement found; flag set to unspecified.' and cite the nearest relevant line(s)."
    }
  },
  "cookie_practices_mentioned": true/false/"unspecified",
  "cookie_practices_excerpt": "string (<= 600 chars). If unspecified, say: 'No explicit mention found.'",
  "mentions_ccpa": true/false/"unspecified",
  "mentions_ccpa_excerpt": "string (<= 600 chars). If unspecified, say: 'No explicit mention found.'",
  "contacts": {
    "opt_out": {
      "form_urls": ["..."],
      "emails": ["..."],
      "phone_numbers": ["..."],
      "other_urls": ["..."]
    },
    "access_or_delete": {
      "form_urls": ["..."],
      "emails": ["..."],
      "phone_numbers": ["..."],
      "other_urls": ["..."]
    }
  },
  "contacts_flat": [
    {
      "type": "form_url|email|phone|other_url",
      "for": "opt_out|access|delete|access_delete|general_privacy",
      "value": "string",
      "label": "string or empty"
    }
  ]
}

## Section A - Policy Scope Categorization
- online_data_practices: Briefly summarize policy text on web/app tracking and collection (e.g., cookies, pixels, SDKs, analytics, ad tech, web forms, account signup). If not present, return "Not mentioned."
- offline_data_practices: Briefly summarize policy text on in-store collection, call centers, paper forms, events, or other in-person channels. If not present, return "Not mentioned."

## Section B - Policy Quality Assessment (Scored 0-3)
Score using only what the policy explicitly states.

### 1) completeness_score (0-3)
Assess whether the policy mentions and explains consumer rights central to CCPA and how to exercise them.
- What to look for: Access/know; deletion; opt-out of sale/share/targeted advertising; (nice-to-have but not required: correction, portability, non-discrimination, appeal).
- Scoring guide:
  - 0: No meaningful mention of rights or how to exercise them.
  - 1: Vague rights mention, minimal specifics, unclear processes.
  - 2: Rights identified with at least some concrete instructions (e.g., link/email/form/phone), but gaps remain (e.g., missing one core right or unclear verification).
  - 3: Rights clearly listed (access/know, delete, opt-out of sale/share/targeted advertising), with actionable mechanisms (webform/email/phone link) and basics on verification, timelines, and scope.

### 2) usability_score (0-3)
Evaluate how actionable the mechanisms are.
- What to look for: Dedicated links/buttons, DSAR portals, emails, phone numbers, clear steps, eligibility/verification guidance, response timelines, required information.
- Scoring guide:
  - 0: No method to act; purely informational.
  - 1: A method exists but is hard to find or ambiguous (e.g., generic support email, broken/missing link).
  - 2: Clear method(s) with some guidance; minor friction or missing details.
  - 3: Multiple clear methods + step-by-step clarity (verification, timelines, exceptions, appeal) and unambiguous labels (e.g., "Do Not Sell or Share My Personal Information").

### 3) accuracy_score (0-3)
Evaluate legal/terminology correctness and internal consistency.
- What to look for: Correct use of terms (e.g., "sale" includes broad consideration; "share" for targeted advertising; "verified consumer"; "sensitive personal information" limits; "household"/"personal information" definitions), avoidance of outdated/incorrect claims, correct opt-out signal references (e.g., GPC).
- Scoring guide:
  - 0: Clearly incorrect or misleading legal statements.
  - 1: Several inaccuracies or outdated framing; ambiguous on critical terms.
  - 2: Mostly accurate with minor issues/ambiguities.
  - 3: Accurate, current, and aligned with CCPA/CPRA language; exceptions and scope are described correctly.

### disclosure_map (booleans)
Mark true only if the policy explicitly discloses the item.
- data_collected: Lists categories collected (e.g., identifiers, geolocation, internet activity).
- data_shared: States data is shared with third parties (including for targeted advertising/cross-context behavioral advertising).
- purpose_of_collection: Explains why data is collected (e.g., service provision, security, marketing, analytics).
- retention_period: States a duration/criteria for retention.
- right_to_access: Explains the right to know/access and how to exercise it.
- right_to_delete: Explains the right to delete and how to exercise it; may list exceptions.
- opt_out: Provides a mechanism to opt out of sale or share/targeted advertising (e.g., DNSMPI page, "Your Privacy Choices" link, opt-out signal).

## Section C - Semantic Behavior Claims (True/False/"unspecified")
Use true/false ONLY if the policy explicitly supports the claim. Otherwise use "unspecified".
For every behavioral flag, provide a non-empty justification:
- If explicit: quote/paraphrase the line(s) and cite the section title if available.
- If not explicit: write, "No explicit statement found; flag set to unspecified." and include the nearest relevant line(s) that discuss cookies/tracking/opt-out signals/third-party sharing, if any.

## Section D - Cookie & CCPA Mentions + Contact Extraction
Strict rules:
- Use true/false ONLY if explicit. Otherwise use "unspecified".
- cookie_practices_mentioned = true if the policy discusses cookies (e.g., cookies, pixels, SDKs, analytics cookies) at all, regardless of consent steps.
- mentions_ccpa = true if the policy explicitly mentions "CCPA", "CPRA", "California Consumer Privacy Act", "California privacy rights" (in a legal-rights sense).
- Extract concrete contact methods for (a) opt-out of sale/share/targeted advertising, and (b) access/delete requests (right to know/delete). Put each into the appropriate bucket(s).
- Provide a flattened list in contacts_flat with the intended purpose (for) and a guessed label if present (e.g., "Your Privacy Choices", "Do Not Sell or Share").
- If no contacts are found for a category, return empty arrays.

## Important Rules
- Base everything only on the provided policy text.
- Do not assume. If unclear or absent, use false (for disclosure_map booleans) or "unspecified" (for behavioral flags).
- Keep summaries concise, specific, and neutral.
- Return only the JSON object--no prose before or after.

================== BEGIN POLICY TEXT ==================
{policy_text}
=================== END POLICY TEXT ===================
""".strip()
\end{lstlisting}
\label{lst:full-llm-prompt}
\clearpage

%% file: tables/accessible-websites-categories.tex
 \begin{table}[!t]
    \centering
    \footnotesize
    \caption{\textbf{Website Categorization by CCPA Applicability and Organizational Type (998 Analyzed Websites).}}
    \label{tab:accessible-websites-categories}
    \begin{tabular}{l  p{4cm}  r}
    \toprule
    \textbf{Category} & \textbf{Description} & \textbf{Count} \\
    \midrule
    \cellcolor{gray!20}Government & \cellcolor{gray!20}Public sector domains & \cellcolor{gray!20}30\\
    Non-Profit & U.S.-based nonprofit organizations & 47 \\
    \cellcolor{gray!20}Revenue Not Sufficient & \cellcolor{gray!20}For-profits below CCPA threshold & \cellcolor{gray!20}319\\
    Subject to CCPA & For-profits meeting CCPA applicability criteria & 602 \\
    \midrule
    \textbf{Total} & & \textbf{998} \\
    \bottomrule
    \end{tabular}
  \end{table}

%% file: figures/new_figures/consent_flow_cookies_ccpa.tex
 \begin{figure}[!t]
   \centering
\includegraphics[width=0.95\linewidth]{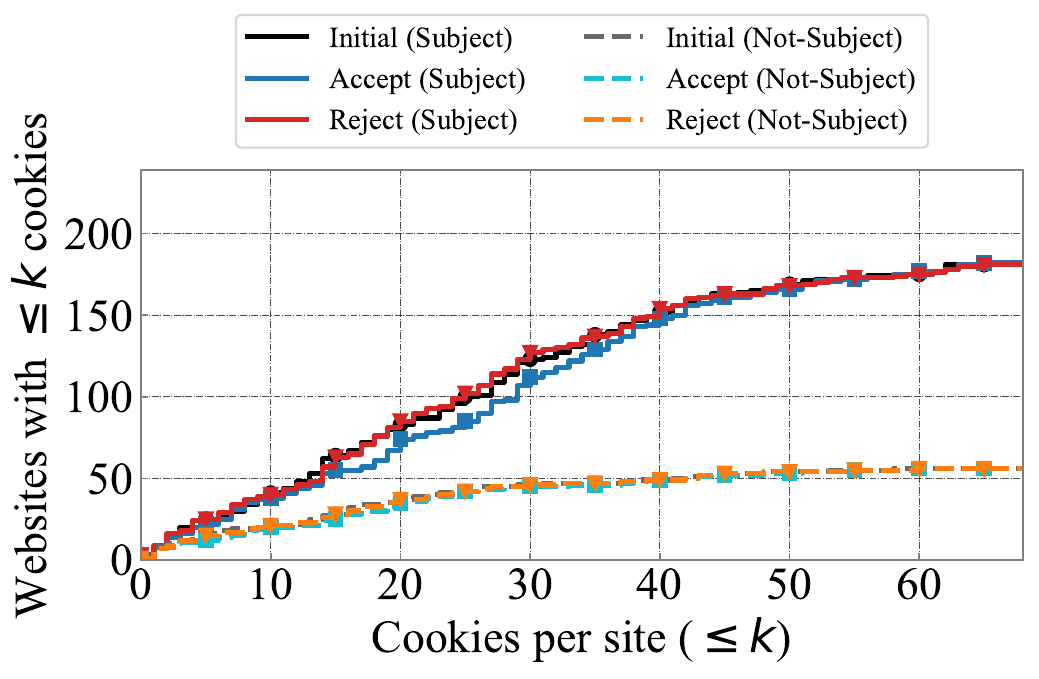}
 \caption{Distribution of per-website cookie counts for CCPA-subject websites (solid lines, N=183) and CCPA-not-subject websites (dashed lines, N=56) across consent stages (Initial, Accept, Reject).}
\label{fig:consent-flow-cookies-ccpa}
\end{figure}

%% file: tables/cookie-attributes-party.tex
  \begin{table}[!t]
      \centering
      \footnotesize
      \caption{Adoption of cookie security attributes (Default browsing configuration).}
      \label{tab:cookie-attributes-party}
      \begin{tabular}{p{1.4cm}
                      >{\raggedright\arraybackslash}p{0.7cm}
                      >{\raggedright\arraybackslash}p{0.9cm}
                      >{\raggedright\arraybackslash}p{0.9cm}
                      >{\raggedright\arraybackslash}p{0.9cm}
                      >{\raggedright\arraybackslash}p{0.8cm}}
      \toprule
      Cookie Type & N & Secure & HttpOnly & SameSite & Session \\
      \midrule
      Targeting & 6,392 & \gradientcell{20.2}{1}{100}{red}{green}{40}\% & \gradientcell{1.4}{1}{100}{red}{green}{40}\% &
  \gradientcell{20.7}{1}{100}{red}{green}{40}\% & \gradientcell{39.5}{1}{100}{red}{green}{40}\% \\
      \rowcolor{gray!20}Performance & 3,464 & \gradientcell{23.1}{1}{100}{red}{green}{40}\% & \gradientcell{2.9}{1}{100}{red}{green}{40}\% &
  \gradientcell{15.5}{1}{100}{red}{green}{40}\% & \gradientcell{14.8}{1}{100}{red}{green}{40}\% \\
      Functional & 2,094 & \gradientcell{51.9}{1}{100}{red}{green}{40}\% & \gradientcell{18.2}{1}{100}{red}{green}{40}\% &
  \gradientcell{24.4}{1}{100}{red}{green}{40}\% & \gradientcell{19.0}{1}{100}{red}{green}{40}\% \\
      \rowcolor{gray!20}Necessary & 1,552 & \gradientcell{50.8}{1}{100}{red}{green}{40}\% & \gradientcell{11.1}{1}{100}{red}{green}{40}\% &
  \gradientcell{27.5}{1}{100}{red}{green}{40}\% & \gradientcell{33.0}{1}{100}{red}{green}{40}\% \\
      Unknown & 5,617 & \gradientcell{16.0}{1}{100}{red}{green}{40}\% & \gradientcell{4.1}{1}{100}{red}{green}{40}\% &
  \gradientcell{5.2}{1}{100}{red}{green}{40}\% & \gradientcell{75.7}{1}{100}{red}{green}{40}\% \\
      \midrule
      \textbf{Total} & \textbf{19,119} & \textbf{\gradientcell{25.4}{1}{100}{red}{green}{40}\%} & \textbf{\gradientcell{5.1}{1}{100}{red}{green}{40}\%}
   & \textbf{\gradientcell{16.1}{1}{100}{red}{green}{40}\%} & \textbf{\gradientcell{42.9}{1}{100}{red}{green}{40}\%} \\
      \bottomrule
      \end{tabular}
  \end{table}

%% file: figures/app-cookie-longevity.tex
\begin{figure}[!t]
  \centering
  \includegraphics[width=\linewidth]{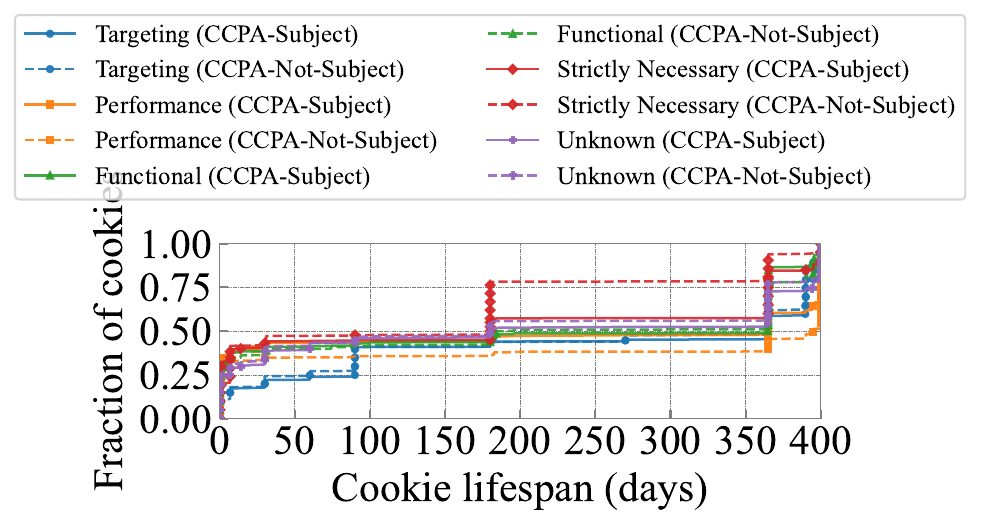}
  \caption{ECDF of per-website non-session cookie longevity by cookie type in the \textit{Default} configuration. Solid lines represent \subject websites and dashed lines represent \notsubject websites. Markers denote sample points.}
  \label{fig:app-cookie-longevity}
\end{figure}

%% file: figures/appendix-industry-policy.tex
\begin{figure}[!t]
  \centering
  \includegraphics[width=\linewidth]{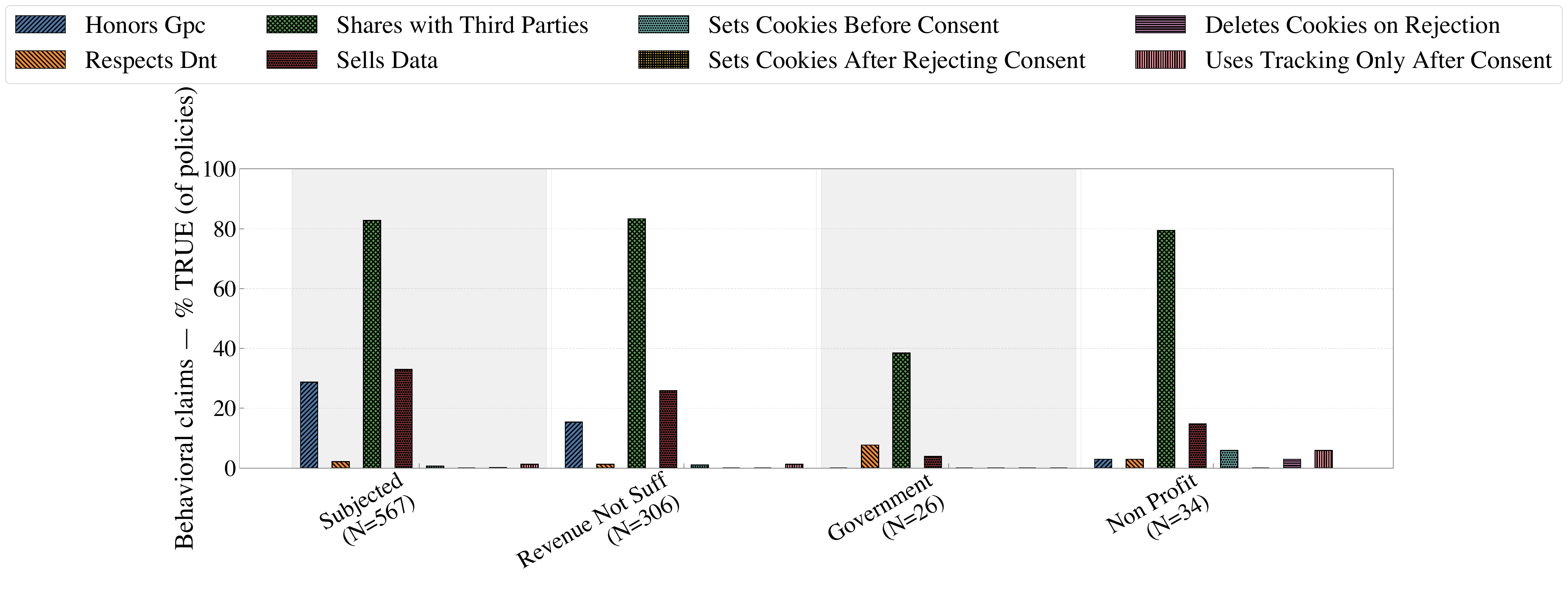}
  \includegraphics[width=\linewidth]{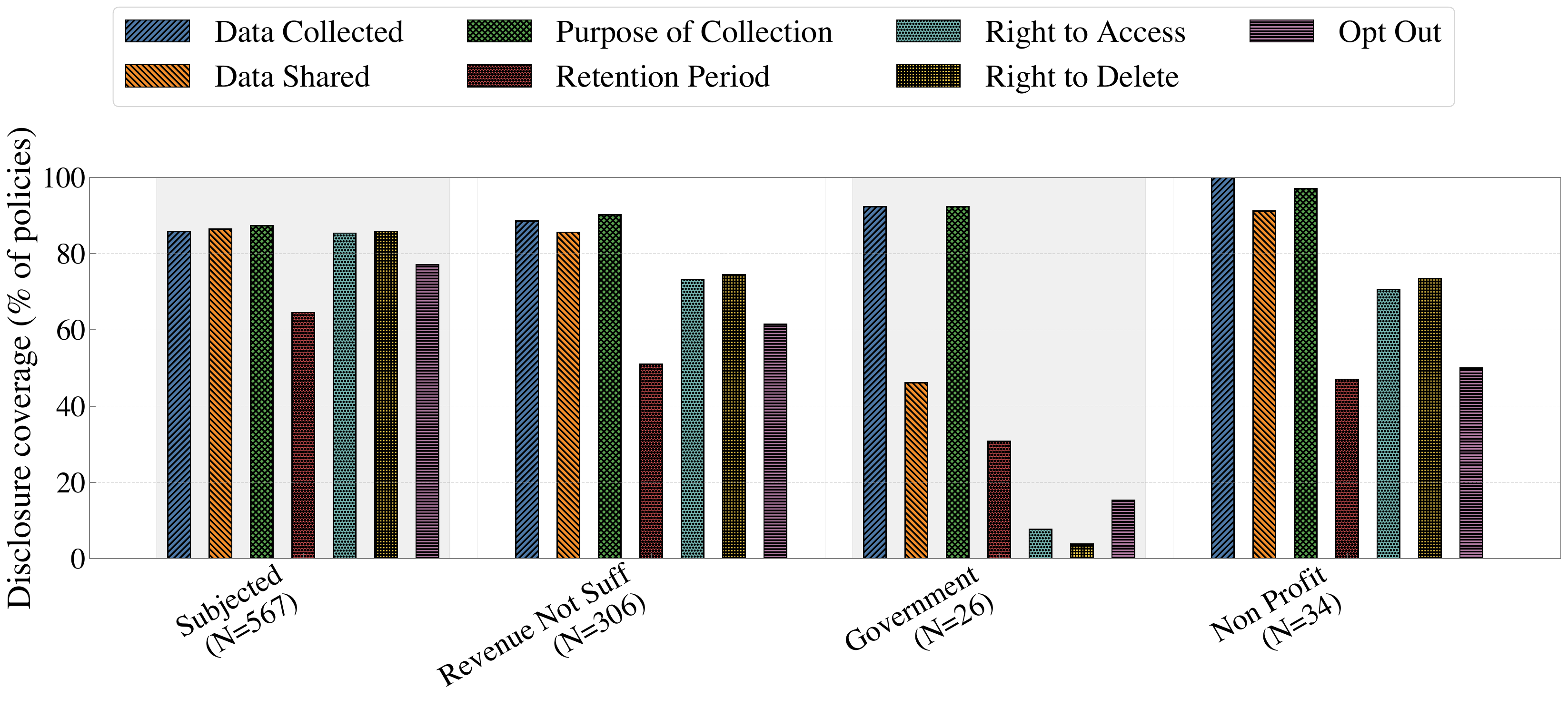}
  \caption{Rates of behavioral claims and disclosure coverage by industry sector. We observe that Government and non-profit websites tend to have lower percentages of both claims and disclosures.}
  \label{fig:appendix-industry-policy}
\end{figure}

%% file: tables/inner-page.tex
\begin{table}[t]
\centering
\footnotesize
\caption{\textbf{Homepage vs.\ internal-page third-party Targeting cookies, by privacy configuration.} Negative paired differences indicate internal pages set fewer third-party Targeting cookies than the homepage. ``Home captures'' is the mean fraction of a \emph{typical internal page's} distinct third-party Targeting trackers also observed on the homepage; the value in parentheses is the stricter fraction of the \emph{union} across all internal pages. Counts reflect the paired homepage and internal-page samples available under each configuration.}
\label{app:innerpage-tab}
\begin{tabular}{p{1cm}rccr}
\toprule
\textbf{Config} & \textbf{Sites} & \textbf{Paired diff (95\% CI)} & \textbf{Also in Home} & \textbf{Add none} \\
\midrule
Default          & 769 & $-0.55$ $[-0.75,-0.36]$ & 87\% (80\%) & 74\% \\
\rowcolor{gray!20}GPC              & 778 & $-0.34$ $[-0.49,-0.21]$ & 87\% (78\%) & 77\% \\
DNT              & 755 & $-0.68$ $[-0.88,-0.48]$ & 89\% (83\%) & 77\% \\
\rowcolor{gray!20}Block 3P         & 756 & $-0.61$ $[-0.80,-0.42]$ & 90\% (83\%) & 78\% \\
uBlock           & 850 & $-0.03$ $[-0.08,\phantom{-}0.02]$ & 87\% (78\%) & 96\% \\
\rowcolor{gray!20}Consent-O-Matic  & 765 & $-0.58$ $[-0.78,-0.40]$ & 89\% (82\%) & 76\% \\
\bottomrule
\end{tabular}
\end{table}

%% file: tables/firm-size.tex
\begin{table}[!t]
\centering
\footnotesize
\caption{Disclosure differences persist after controlling for firm size and ad-revenue reliance.
\textbf{(A)} Logistic-regression coefficients (log-odds) for each disclosure on CCPA subjectivity,
adjusting for company annual revenue ($\log_{10}$) and ad-revenue reliance (each site's third-party
Targeting tracking intensity); $N=816$ sites with revenue data. \textbf{(B)} Disclosure rates among
large organizations only (annual revenue $\geq$\$25M): CCPA-subject firms vs.\ large but exempt
organizations (nonprofits/government); $N=563$ (528 subject, 35 exempt).
$^{*}p<0.05$, $^{**}p<0.01$, $^{***}p<0.001$.}
\label{tab:size-control}
\begin{tabular}{l ccc}
\toprule
\multicolumn{4}{@{}l}{\textbf{(A) Logistic regression (log-odds)}} \\
\textbf{Disclosure} & \textbf{CCPA Subject} & \textbf{$\log_{10}$Revenue} & \textbf{Ad-reliance} \\
\midrule
Opt-out & $+0.98^{***}$ & $-0.05$ & $+0.91^{***}$ \\
\rowcolor{gray!20}Access  & $+1.14^{***}$ & $-0.16$ & $+0.73^{***}$ \\
Delete  & $+1.15^{***}$ & $-0.18^{*}$ & $+0.73^{**}$ \\
\rowcolor{gray!20}GPC     & $+0.55^{*}$   & $+0.17^{*}$ & $+0.74^{***}$ \\
\midrule
\multicolumn{4}{@{}l}{\textbf{(B) Large organizations only ($\geq$\$25M revenue)}} \\
\textbf{Disclosure} & \textbf{Subject} & \textbf{Exempt} & \textbf{$p$} \\
\midrule
Opt-out & 77\% & 46\% & $<$0.001 \\
\rowcolor{gray!20}Access  & 85\% & 60\% & $<$0.001 \\
Delete  & 85\% & 60\% & $<$0.001 \\
\rowcolor{gray!20}GPC     & 28\% & 0\%  & $<$0.001 \\
\bottomrule
\end{tabular}
\end{table}

%% file: tables/corrections.tex
\begin{table*}[!t]
\centering
\footnotesize
\caption{Raw and multiple-testing--adjusted $p$-values for all 25 reported Subject vs.\ Not-Subject tests (16 policy, 9 cookie), treated as a single family. BH: Benjamini--Hochberg adjusted $p$-value (controls the false-discovery rate, FDR); Holm: Holm--Bonferroni adjusted $p$-value (controls the family-wise error rate, FWER), following O'Connor \etal~\cite{o2021clear}. The last two columns mark significance at $\alpha=0.05$ after adjustment; we base our conclusions on the Holm column.}
\label{app:mht}
\begin{tabular}{@{}llrrrcc@{}}\toprule
Family & Test & $p$-value & BH adj.\ $p$ & Holm adj.\ $p$ & BH$^{*}$ & Holm$^{*}$ \\ \midrule
policy & rubric\_usability & $3.6\times10^{-19}$ & $9.0\times10^{-18}$ & $9.0\times10^{-18}$ & \checkmark & \checkmark \\
\rowcolor{gray!20}policy & claim\_m\_offline & $1.2\times10^{-17}$ & $1.5\times10^{-16}$ & $2.8\times10^{-16}$ & \checkmark & \checkmark \\
policy & rubric\_completeness & $1.2\times10^{-14}$ & $9.7\times10^{-14}$ & $2.7\times10^{-13}$ & \checkmark & \checkmark \\
\rowcolor{gray!20}cookie & 3p\_Performance & $5.8\times10^{-14}$ & $3.6\times10^{-13}$ & $1.3\times10^{-12}$ & \checkmark & \checkmark \\
policy & rubric\_accuracy & $1.2\times10^{-11}$ & $6.2\times10^{-11}$ & $2.6\times10^{-10}$ & \checkmark & \checkmark \\
\rowcolor{gray!20}policy & disc\_opt\_out & $1.1\times10^{-10}$ & $4.6\times10^{-10}$ & $2.2\times10^{-9}$ & \checkmark & \checkmark \\
policy & disc\_right\_to\_access & $5.3\times10^{-10}$ & $1.9\times10^{-9}$ & $1.0\times10^{-8}$ & \checkmark & \checkmark \\
\rowcolor{gray!20}policy & disc\_right\_to\_delete & $1.2\times10^{-9}$ & $3.7\times10^{-9}$ & $2.1\times10^{-8}$ & \checkmark & \checkmark \\
policy & claim\_honors\_gpc & $2.5\times10^{-8}$ & $6.9\times10^{-8}$ & $4.3\times10^{-7}$ & \checkmark & \checkmark \\
\rowcolor{gray!20}cookie & count\_Functional & $1.4\times10^{-7}$ & $3.6\times10^{-7}$ & $2.3\times10^{-6}$ & \checkmark & \checkmark \\
cookie & 3p\_Functional & $3.1\times10^{-7}$ & $7.1\times10^{-7}$ & $4.7\times10^{-6}$ & \checkmark & \checkmark \\
\rowcolor{gray!20}cookie & 3p\_Targeting & $2.0\times10^{-6}$ & $4.1\times10^{-6}$ & $2.7\times10^{-5}$ & \checkmark & \checkmark \\
policy & disc\_retention\_period & $3.3\times10^{-6}$ & $6.3\times10^{-6}$ & $4.3\times10^{-5}$ & \checkmark & \checkmark \\
\rowcolor{gray!20}policy & claim\_sells\_data & 0.001 & 0.002 & 0.016 & \checkmark & \checkmark \\
cookie & count\_Necessary & 0.006 & 0.010 & 0.066 & \checkmark & -- \\
\rowcolor{gray!20}cookie & 3p\_Necessary & 0.035 & 0.055 & 0.352 & -- & -- \\
cookie & count\_Performance & 0.037 & 0.055 & 0.352 & -- & -- \\
\rowcolor{gray!20}policy & disc\_data\_collected & 0.072 & 0.100 & 0.573 & -- & -- \\
policy & disc\_purpose\_of\_collection & 0.082 & 0.108 & 0.576 & -- & -- \\
\rowcolor{gray!20}cookie & 3p\_overall & 0.145 & 0.176 & 0.869 & -- & -- \\
policy & claim\_m\_online & 0.148 & 0.176 & 0.869 & -- & -- \\
\rowcolor{gray!20}policy & disc\_data\_shared & 0.195 & 0.221 & 0.869 & -- & -- \\
policy & claim\_shares\_3p & 0.259 & 0.282 & 0.869 & -- & -- \\
\rowcolor{gray!20}cookie & count\_Targeting & 0.562 & 0.585 & 1.000 & -- & -- \\
policy & claim\_respects\_dnt & 0.830 & 0.830 & 1.000 & -- & -- \\
\bottomrule
\end{tabular}
\end{table*}

%% file: tables/legal-basis.tex
  \begin{table*}[!t]
  \footnotesize
  \centering
  \caption{Mapping of LLM-Prompt JSON output Fields to specific CCPA/CPRA Provisions}
  \label{tab:ccpa_legal_mapping}
  \begin{tabular}{p{3.4cm} p{2cm} p{11.3cm}}
  \toprule
  \textbf{Field} & \textbf{Legal Reference (Cal. Civ. Code)} & \textbf{Rationale} \\
  \midrule
  \rowcolor{gray!20}categories\allowbreak\_of\allowbreak\_personal\allowbreak\_information\allowbreak\_collected & \S1798.130\allowbreak(a)\allowbreak(5)\allowbreak(B)\allowbreak(i) & Requires
  disclosure in the privacy policy of categories of personal information collected about consumers. \\
  categories\allowbreak\_of\allowbreak\_sensitive\allowbreak\_personal\allowbreak\_information & \S1798.100\allowbreak(a)\allowbreak(2), \S1798.100\allowbreak(e), \S1798.121 & If sensitive personal
  information is collected, businesses must disclose the categories and purposes for collection, and whether it is sold or shared. CPRA establishes
  additional disclosure and limitation rights for sensitive personal information. Regulated entities must also ensure reasonable security of sensitive personal information. \\
  \rowcolor{gray!20}purpose\allowbreak\_of\allowbreak\_collection & \S1798.130\allowbreak(a)\allowbreak(5)\allowbreak(B)\allowbreak(iii) & Businesses must state the business or commercial purpose
   for collecting or sharing personal information. \\
  categories\allowbreak\_of\allowbreak\_sources & \S1798.130\allowbreak(a)\allowbreak(5)\allowbreak(B)\allowbreak(ii) & Requires disclosure of categories of sources from which personal
  information is collected. \\
  \rowcolor{gray!20}categories\allowbreak\_of\allowbreak\_third\allowbreak\_parties & \S1798.130\allowbreak(a)\allowbreak(5)\allowbreak(B)\allowbreak(iv) & Requires disclosure of categories of
  third parties with whom personal information is shared. \\
  sale\allowbreak\_of\allowbreak\_personal\allowbreak\_information & \S1798.130\allowbreak(a)\allowbreak(5)\allowbreak(C)\allowbreak(i) & Businesses must disclose categories of personal
  information sold, or affirmatively state if no sale occurred in the past 12 months. \\
  \rowcolor{gray!20}sharing\allowbreak\_of\allowbreak\_personal\allowbreak\_information & \S1798.130\allowbreak(a)\allowbreak(5)\allowbreak(C)\allowbreak(i) & Under CPRA, ``sharing'' for
  cross-context behavioral advertising must be disclosed in the same manner as ``sales.'' \\
  disclosure\allowbreak\_for\allowbreak\_business\allowbreak\_purpose & \S1798.130\allowbreak(a)\allowbreak(5)\allowbreak(C)\allowbreak(ii) & Requires disclosure of categories of personal
  information disclosed for a business purpose in the past 12 months. \\
  \rowcolor{gray!20}right\allowbreak\_to\allowbreak\_know & \S1798.110 & Consumers have a right to know what personal information is collected, used,
  or disclosed. Businesses must disclose this right and provide mechanisms to exercise it. \\
  right\allowbreak\_to\allowbreak\_delete & \S1798.105 & Grants consumers the right to request deletion of their personal information (subject to limited exceptions such as completing a transaction or complying with legal obligations). Businesses must disclose this right and the process to exercise it, and respond within 45 days.
 \\
  \rowcolor{gray!20}right\allowbreak\_to\allowbreak\_correct & \S1798.106 & CPRA establishes the right to request correction of inaccurate personal
  information; businesses must disclose this right and provide mechanisms to exercise it. \\
  right\allowbreak\_to\allowbreak\_opt\allowbreak\_out & \S1798.120 & Consumers have the right to opt out of the sale or sharing of personal
  information. Businesses must provide clear notice of this right and mechanisms to exercise it. \\
  \rowcolor{gray!20}right\allowbreak\_to\allowbreak\_limit\allowbreak\_sensitive\allowbreak\_personal\allowbreak\_information & \S1798.121\allowbreak(a) &
  Consumers may limit use and disclosure of sensitive personal information to purposes reasonably necessary to perform services or provide goods;
  businesses must enable this right through a ``Limit the Use of My Sensitive Personal Information'' link. \\
  non\allowbreak\_discrimination & \S1798.125\allowbreak(a) & Prohibits discrimination (e.g., denial of goods/services or unequal pricing) against consumers who
  exercise their CCPA rights, though businesses may charge different prices if reasonably related to the value provided by consumer data. \\
  \rowcolor{gray!20}financial\allowbreak\_incentives & \S1798.125\allowbreak(b) & Businesses may offer financial incentives for collection, sale, or retention of
  personal information; must provide notice of material terms and allow consumers to opt in (with revocable consent) or opt out. Cannot be unjust,
  unreasonable, coercive, or usurious. \\
  authorized\allowbreak\_agent & \S1798.135\allowbreak(e) & Consumers may designate an authorized agent to exercise rights on their behalf; businesses must honor
  such requests pursuant to Attorney General regulations. \\
  \rowcolor{gray!20}methods\allowbreak\_for\allowbreak\_requests & \S1798.130\allowbreak(a)\allowbreak(1) & Requires businesses to provide two or more designated methods
  (e.g., toll-free number and website/webform) to submit consumer rights requests. \\
  data\allowbreak\_retention & \S1798.100\allowbreak(a)\allowbreak(3) & Requires disclosure of retention period for each category of personal information (including
  sensitive personal information), or criteria used to determine it. Personal information must not be retained longer than reasonably necessary for
  disclosed purposes. \\
  \rowcolor{gray!20}policy\allowbreak\_last\allowbreak\_updated & \S1798.130\allowbreak(a)\allowbreak(5) & Privacy policies must be updated at least once every 12 months to
  reflect current practices and disclosures. \\
  \bottomrule
  \end{tabular}
  \end{table*}